    \documentstyle[12pt]{article}
    
    \newcommand{\acton}[2]
    {\stackrel{\rule{.1mm}{1.95mm}\rule[1.85mm]{#1}{.1mm}
    \hspace{-.6mm}\downarrow}{#2}}
    \newcommand{\tr}{\mbox{tr}}
    \renewcommand{\theequation}{\arabic{section}.\arabic{equation}}
    
\begin{document}
      \author{}
   \title{The one-loop form factors in the effective action,\\
    and \\
    production of coherent gravitons from the vacuum.}
    \date{}
    \maketitle
    \begin{center}
   {\large A.G.Mirzabekian} \\
   Lebedev Research Center in Physics, Leninskii
   Prospect 53,\\ Moscow 117924, Russia\\
{\large G.A.Vilkovisky} \\
   Lebedev Research Center in Physics, Leninskii
   Prospect 53,\\ Moscow 117924, Russia\\
   and \\
   Istituto Nazionale
   di Fisica Nucleare, Sezione di Napoli,\\
   Pad. 20 Mostra d'Oltremare,Napoli 80125, Italia
   \end{center}
 \newpage
\begin{abstract}
We present the solution of the problem of the $1/\Box, \Box \to 0,$
asymptotic terms discovered in the one-loop form factors of the
gravitational effective action. Owing to certain constraints among their
coefficients, which we establish, these terms cancel in the vacuum stress
tensor and do not
 violate the asymptotic flatness of the expectation value of the metric.
They reappear, however, in the Riemann tensor of this metric and
stand for a new effect: a radiation of gravitational waves induced
by the vacuum stress. This coherent radiation caused by the
backreaction adds to the noncoherent radiation caused by the pair
creation in the case where the initial state provides the vacuum
stress tensor with a quadrupole moment.
\end{abstract}
\newpage
\section{Introduction}
$\;\;$

We consider the expectation-value equations for the gravitational
field in an in-state \footnote{
The gravitational collapse problem was first considered in this
setting in refs. [1-3].}.
 The model-independent, or phenomenological approach [4-7]
makes it possible to write down the general form of these equations
in terms of the form factors in the vacuum action. The form factors
are to be calculated from a given dynamical model. However, for
obtaining  predictions of various models, the expectation-value
equations should be first analysed with arbitrary form factors in order
to relate the properties of the form factors to the important properties
of the solution [4].

This analysis has thus far been limited to the behaviour of the vacuum
stress tensor $T^{\mu\nu}_{vac}$ at infinity [6,7]. It was shown [6]
that the requirement of asymptotic flatness of the solution imposes
restrictions on the behaviours of the form factors at small values of their
arguments. Namely, these behaviours in one (each) of the arguments
with the others fixed should be $w(0)\log (-\Box) + O(1), \Box \to -0.$ The
coefficients $w(0)$ of the $\log (-\Box), \Box \to -0$ behaviours
(the spectral weights at zero spectral mass) determine the rate of the vacuum
radiation through the future null infinity $({\cal I}^+).$

However, at each given order  in the curvature, only certain combinations
of the form factors must behave in this way. The condition of finiteness of
$T^{\mu\nu}_{vac}$ at ${\cal I}^+$ which brought to the result above
leaves some arbitrariness in the asymptotic behaviours of the individual
form factors in the basis decomposition of the action. It is this arbitrariness
that allows for the existence of the effect discussed in the present paper.

The present paper deals with the problem which appeared when the
field-theoretic form factors were calculated. It is worth noting that, as
emphasized in [6], the loop expansion of field theory has a domain of
validity. It is near null infinity where the results of this expansion are
valid and can be used to calculate the energy of the vacuum radiation.
The ultraviolet divergent terms which appear in the expectation-value
equations when $T^{\mu\nu}_{vac}$ is expanded in loops are local
and vanish at infinity; only nonlocal terms survive, and these are
unambiguous. Technically, the loop expanded form factors are reliable
in the limit of small $\Box$ arguments up to terms
$O({\Box}^0), \Box \to -0.$

The field-theoretic form factors have been calculated in the one-loop
approximation for a generic quantum field model in refs. [8-15]. Their
asymptotic behaviours at $\Box \to -0$ up to terms $O({\Box}^0),$
which are of interest for the above-mentioned reason, are presented in
ref. [15], and these behaviours offer a problem. While the second-order
form factors behave as expected: they are $w(0) \log (-\Box) + O(1)$
for $\Box \to -0,$ the third-order form factors contain also the asymptotic
terms $1/\Box, \Box \to -0$ which apparently violate the asymptotic flatness
of the solution [6,15].

Since the third-order form factors are functions of three $\Box$ arguments,
their behaviours in one of the arguments with the two others fixed cannot
be predicted on dimensional grounds. The $1/\Box$ asymptotic terms in the
form factors appear as a result of an explicit calculation of loops [12,15]. In
the present paper we propose an explanation of this result as well as of the
following remarkable fact [12] which one can  establish by a direct
inspection of the expressions in [15]. The inspection shows that the alarming
$1/\Box$ terms appear only in the curvature invariants containing the
gravitational field strength and act {\it selectively only} on the Ricci
curvature. The matter field strengths contained in the commutator and
potential curvatures (see [15]) remain unaffected by these terms.
Thus the presence of the $1/\Box$ terms breaks the democracy of massless
vacuum particles; gravitons appear to be distinguished.

Since the same vacuum action describes also the transition amplitudes
between in- and out- states [8], the problem of the  $1/\Box$ terms
appears also in scattering theory where these terms
either signal an infrared divergence of the on-shell amplitudes with
gravitons or, in the favourable case, stand for some inelastic process.
In the language of expectation values, the former case corresponds to a
breakdown of the asymptotic flatness. That the situation is not hopeless
is seen from the fact that the form factors are not quite the vertices; they
are coefficients of the curvatures rather than the field disturbances.
As compared to the field disturbances, the curvatures contain extra
derivatives which, in the favourable case, may cancel the $1/\Box$ factors
in the on-shell amplitudes.

Below we present the solution of the problem as it appears in
expectation-value theory. As mentioned above, only certain combinations
of the form factors should behave like $\log (-\Box),\Box \to -0$ to ensure
finiteness of $T^{\mu\nu}_{vac}$ at ${\cal I}^+.$
We show that the $1/\Box$ terms precisely cancel in these combinations
leaving indeed the $\log (-\Box)$ behaviour as the leading one. The
cancellation occurs owing to certain constraints among the coefficients
of the $1/\Box$ terms which we establish by an analysis of the asymptotic
flatness and next check with the explicit expressions in [15].
The fulfilment of these constraints is by itself a powerful check on the
results in [15] apart from the checks that have already been carried out
in [12]. Thus we prove that the $1/\Box, \Box \to -0$ terms discovered
in [12,15] do not violate the asymptotic flatness of the solution of the
expectation-value equations. The proof is given for a generic quantum
field model for which the results in [15] are obtained and which is
characterized by a set of field strengths consisting of the Ricci, commutator,
and potential curvatures. Such a general proof is possible owing to the
above-mentioned fact that the $1/\Box$ operators in the asymptotic
expressions for the form factors act only on the Ricci curvature.
Therefore, for the consideration of the leading asymptotic terms of the
equations
at ${\cal I}^+,$ one does not need to know the variational derivatives
of the commutator and potential curvatures with respect to the metric.

Next, we reveal the significance of the $1/\Box, \Box \to -0$ terms in the
vacuum form factors. These terms vanish in the energy-momentum tensor but , as
we show,
 they reappear in the Riemann tensor of the solution and determine its leading
$O(1/r)$ behaviour at ${\cal I}^+.$ The coefficient of this behaviour is known
 to give the energy of the outgoing gravitational waves. Thus the $1/\Box, \Box
\to -0$ terms of the vacuum form factors discovered in [12,15] stand for
a new effect: a generation of the gravitational waves from the vacuum.
This is an effect of the backreaction of the vacuum stress on the metric.
All massless particles including gravitons [16] contribute to
 $T^{\mu\nu}_{vac}$ and are radiated through the future null infinity
by the quantum mechanism of pair creation. The energy of this component
 of radiation is determined by the $\log (-\Box), \Box \to -0$ terms of the
vacuum form factors. On top of this, $T^{\mu\nu}_{vac}$ as a whole acts
as a source of the gravitational field and causes a secondary radiation of
gravitons. This component of radiation has the shape of a classical wave
 but with a quantum amplitude, and its energy is determined by the
$1/\Box , \Box \to -0 $ terms of the vacuum form factors.

An important difference between the two cases is that the
gravitational-wave component will be nonvanishing only if the initial
state has a sufficient asymmetry to provide the vacuum stress with
a quadrupole moment whereas the contribution of gravitons in
$T^{\mu\nu}_{vac}|_{{\cal I}^+}$ is present even in a spherically
symmetric in-state because the out-states in which these gravitons
appear at ${\cal I}^+$ are squeezed vacuum states rather than
coherent states (see [17,18] and references therein).

The paper is organized as follows. In sec. 2  we briefly review the
structure of the expectation-value equations and present a new
expression for the solution of the Bianchi identities to second
order in the curvature. This expression simplifies obtaining the
news functions of the gravitational waves. Sec. 3 contains an
analysis of contributions to $T^{\mu\nu}_{vac}$ at ${\cal I}^+.$
This analysis is not complete but sufficient for obtaining the
asymptotic flatness constraints and for calculating
$T^{\mu\nu}_{vac}|_{{\cal I}^+}$ to second order in the
curvature. We point out an important distinction of this
calculation from the one in two dimensions [3], and also make
some first step in giving the expectation-value equations a
closed form.
 Sec. 4 contains the derivation of the asymptotic flatness
constraints and the proof that they are satisfied by the
 one-loop form factors. In sec. 5 we propose a new method
for calculating the energy of the gravitational waves and
obtain the contribution of the third-order vacuum form
factors to the news functions. Sec. 6 completes the
calculation of the vacuum news functions in the lowest
nonvanishing approximation. Appendix A contains
reference equations pertaining to the behaviour of
the asymptotically flat metric at null infinity.
Appendix B summarizes the properties of the retarded
 Green function used in the text.
    \setcounter{equation}{0}
        \section{ The expectation-value equations in an in-state}
$\mbox{}$

In the framework of quantum field theory one starts with the assumption
that there exists a quantum state such that the expectation value of the
metric in this state is an asymptotically flat gravitational field. Under
 this assumption one goes to the past null infinity $({\cal I}^-)$ of
the spacetime with the expectation value of the metric and, for all
massless fields, builds the Fock space of states (the in-states) based
 on the standard in-vacuum [19]. The assumed state belongs to this space
\footnote{After the choice of the state has been made, one checks the
 original assumption. Since the causality relationship is set by
 the expectation
value of the metric, this is a self-consistent problem even in its
original setting. That the consistency check is nontrivial is
seen, for example,
 from the fact that with the massive quantum fields one generally
 arrives at a
 contradiction with the asymptotic flatness.}. The choice of
the state determines the initial data at ${\cal I}^-$ for the field's
expectation values and, generally, affects also the form of their dynamical
equations since these equations are nonlocal. There exists an action
which produces the expectation-value equations although the procedure
by which it does so {\it is not} the least-action principle (see below).

It makes sense to choose the initial state in which the matter quanta form
some heavy classically behaved source of the gravitational field, and
gravitons are in a coherent state so that, for the mean metric, they form,
generally, a classical incoming gravitational wave. Here we consider
the case where such a wave is absent. The action for the expectation
value of the metric in such a state can  be taken as the sum
\begin{equation}
     S = S_{vac} + S_{source}
\end{equation}
where $S_{source}$ is the action of a source which moves along a
classical trajectory in the mean metric, and $S_{vac}$ is the action for
the gravitational field in the in-vacuum state.

The action $S_{vac}$ is to be calculated from a given quantum field
model. Within certain approximations (which are not completely
unsatisfactory, see above) this calculation is feasible and, in the
 one-loop approximation, it can be done for a generic field model
[8-15]. However, since the gravitational interaction is universal,
all particles existing in nature contribute to $S_{vac},$ and, at
higher loop orders, all details of their interactions matter.
Therefore, if $S_{vac}$ is to be {\it ultimately} calculated from a model
(of fields or strings or whatever), then this should be the Model and
the Calculation.

In the phenomenological approach of refs. [4-7], the action (2.1) is
viewed as an effective action (in the loose sense) for the observable
field which should be a part of predictions of any fundamental
dynamical theory. Irrespectively of the nature of this theory, one
assumes the existence of a functional, the action $S_{vac},$
which describes the elastic properties of real vacuum i.e. its
response to the introduction of a gravitationally charged source.
For $S_{vac}$ one writes down the most general expansion in
terms of nonlocal invariants of $N$th order in the curvature. One
has to go explicitly to $N=3$ because third order in the curvature
in the action corresponds to second order in the equations, and it
has been shown that, at first order in the curvature, the flux of
 vacuum energy through ${\cal I}^+$ is pure quantum noise [7].
The full bases of nonlocal invariants of second order  and third
order are built in [5] for a set of field strengths consisting of the Ricci,
commutator and potential curvatures:
\begin{equation}
                    \Re = \left\{ R_{\mu\nu}, {\hat {\cal R}}_{\mu\nu},
                                                 {\hat P} \right\}
\end{equation}
( for the definitions see [15]). The explanation
for the absence of the Riemann
tensor from the basis invariants
can be found in [9,4,5] but it makes sense to repeat it
here since the respective equations will be of use below.

By differentiating and contracting the Bianchi
identities, one obtains the equation
\begin{eqnarray}
 \Box R^{\alpha\beta\mu\nu} =
2 \nabla^{[\mu}\nabla^{\alpha} R^{\nu] \beta}
- 2 \nabla^{[\mu}\nabla^{\beta} R^{\nu] \alpha} -
4 R_{\:.\gamma\:.\:\sigma}^{\alpha\:[\mu} R^{\beta\gamma\nu] \sigma}
+ 2 R^{[\mu}_{\;\:\gamma} R^{\alpha\beta\gamma\nu]} - \nonumber \\
{} - R^{\alpha\beta}_{\ \ \gamma\sigma} R^{\mu\nu\gamma\sigma}
\qquad\qquad
\end{eqnarray}
which can be solved iteratively with respect to the Riemann tensor. In
this equation the Ricci tensor plays the role of a source, and the solution
is fixed by the initial data for the gravitational field at ${\cal I}^-.$
The solution with zero data (no incoming gravitational wave) corresponds
to the in-vacuum state and is expressed in terms of the retarded Green
 function (see appendix B):
\renewcommand{\theequation}{\arabic{section}.4a}
\begin{equation}
R^{\alpha\beta\mu\nu} = \frac{1}{\Box} \Bigl(
4 \nabla^{[\mu}\nabla^{<\alpha} R^{\nu]\beta>} +
O[R^2_{..}] \Bigr) .
\end{equation}
Here and below, $1/\Box$ stands for the retarded Green function,
and both types of brackets $[\: ]$ and $< >$ denote antisymmetrization
of the respective indices. Since the Riemann tensor is expressed in
 this way through the Ricci tensor, the nonlocal invariants with
the Riemann tensor in the vacuum action are redundant.

Below we shall confine ourselves to the case where the flux
components of the Ricci tensor at ${\cal I}^-$ vanish
\footnote{
For the expectation value of the metric this condition restricts
 the choice of the classical source which can be taken, e.g.,
as having a compact spatial support. Then, since the
energy-momentum tensor of the in-vacuum has no incoming
fluxes at ${\cal I}^-,$ the same will be true also of
 $R^{\mu\nu}$ (see eq. (2.18) below). In the general case,
the flux components of $R^{\mu\nu}$ at ${\cal I}^-$ cancel
in the combination $\nabla^{[\mu} \nabla^{<\alpha} R^{\nu ]\beta >}$
appearing in (2.4a) but the derivatives cannot be commuted with
$1/\Box$ for otherwise the action of the retarded Green function
will become ill-defined.}.
In this case the derivatives in (2.4a) can be made external by
 commuting them with the Green function $1/\Box.$ To second
order in the Ricci curvature the solution is then of the form
\renewcommand{\theequation}{\arabic{section}.4b}
\begin{eqnarray}
R^{\alpha\beta\mu\nu}
 &=& \nabla^{[\mu} \nabla^{<\alpha}
\left(\frac{4}{\Box}\right) \Bigl[ R^{\nu] \beta >} +
 \Bigl( \nabla^{\nu ]} \frac{1}{\Box} R^{\gamma\delta} \Bigr)
\Bigl( \nabla^{\beta >} \frac{1}{\Box} R_{\gamma\delta} \Bigr) -
\\
& & {} -   2\Bigl( \nabla_{\gamma} \frac{1}{\Box} R^{\nu ]\delta} \Bigr)
\Bigl( \nabla_{\delta} \frac{1}{\Box} R^{\beta >\gamma} \Bigr)  \Bigr] +
{} \biggl\{ 8g_{\gamma\delta}\Bigl(\frac{1}{\Box}
\nabla^{[\alpha}R^{<\mu\gamma]}\Bigr)\Bigl(\frac{1}{\Box}
\nabla^{[\beta}R^{\nu>\delta]}\Bigr) + \nonumber \\
 & & {} + 8g_{\gamma\delta}\Bigl(\frac{1}{\Box}
\nabla^{[\mu}R^{<\alpha\gamma]}\Bigr)\Bigl(\frac{1}{\Box}
\nabla^{[\nu}R^{\beta>\delta]}\Bigr) -
 2\Bigl(\frac{1}{\Box}
\nabla_{\gamma}R^{<\alpha[\mu}\Bigr)\Bigl(\frac{1}{\Box}
\nabla^{\gamma}R^{\beta>\nu]}\Bigr) + \nonumber \\
 & & {} + 4\Bigl(\frac{1}{\Box}\nabla^{[\mu}\nabla^{\gamma}R^{
\nu]<\alpha}\Bigr)\Bigl(\frac{1}{\Box}R^{\beta>}_{\:\gamma}
\Bigr) + 4\nabla^{[\mu}\Bigl(\Bigl[\frac{1}{\Box}
\nabla^{<\alpha}R^{\beta>\gamma}\Bigr]\Bigl[
\frac{1}{\Box}R^{\nu]}_{\:\gamma}\Bigr]\Bigr)\biggr\} +
  O[R^3_{..}] \nonumber
\end{eqnarray}
(cf. the result in [5,12]). The advantage of making the derivatives
external is in the appearance of the terms in the curly brackets in
(2.4b) which have no overall $1/\Box$ factor. These terms
{\it do not contribute } to the leading asymptotic behaviour of the
Riemann tensor at ${\cal I}^+ .$ On the other hand, the terms
which have the overall $1/\Box$ factor have also the overall
derivatives. This facilitates solving the equation for the news functions
of the gravitational waves ( see sec. 5).
\setcounter{equation}{4}
\renewcommand{\theequation}{\arabic{section}.\arabic{equation}}

To third order in the curvature the vacuum action is of the form
\footnote{We follow the notation of ref. [15] but change the overall
sign of the action as appropriate for the lorentzian signature of the
 metric. We use the signature $(- + + +)$ and the conventions
$ R^{\mu}_{\;\alpha\nu\beta} = \partial_{\nu}
\Gamma^{\mu}_{\alpha\beta} - ... , R_{\alpha\beta} =
R^{\mu}_{\;\alpha\mu\beta} , R = g^{\alpha\beta} R_{\alpha\beta}.$}
\begin{equation}
S_{vac} = S(1) + S(2) + S(3) + O[ \Re^4 ] ,
\end{equation}
\begin{equation}
S(1) = \frac{1}{16\pi} \int dx g^{1/2} R   ,
\end{equation}
\begin{equation}
S(2) = \frac{1}{2{(4\pi)}^2} \int dx g^{1/2} \mbox{tr}
\sum_{i=1}^{5} \gamma_{i} (-\Box_2) \Re_1 \Re_2 (i) ,
\end{equation}
\begin{equation}
S(3) = \frac{1}{2{(4\pi)}^2} \int dx g^{1/2} \mbox{tr}
\sum_{i=1}^{29} \Gamma_{i} (-\Box_1,-\Box_2,-\Box_3)
\Re_1 \Re_2 \Re_3 (i) ,
\end{equation}
\begin{equation}
\Box = g^{\mu\nu} \nabla_{\mu} \nabla_{\nu}
\end{equation}
where  $ \Re_1 \Re_2 (i) $ with $ i=1$ to 5, and
$\Re_1 \Re_2 \Re_3 (i)$ with $i=1$ to 29 are the quadratic and cubic
basis invariants listed in [15]. This list is reduced in comparison to the full
list in [5] because, in the trace of the heat kernel and hence in the
one-loop vacuum action, the invariants linear in the commutator
curvature all but one prove to be absent [12,13].
The only one that is present is number 13 in the list of ref. [15] which
we use here. In low-dimensional manifolds there exist hidden
constraints between nonlocal invariants, reducing the basis. In four
dimensions, the second-order basis is unconstrained, and the only
constraint which exists among the third-order invariants boils down
 to the condition that the completely symmetric part of the form
factor $\Gamma_{28} $ vanishes identically [5,12]. In the
field-theoretic form factors of refs. [12,15] this condition is
explicitly implemented.

The commutator and potential curvatures are functions of the metric
and matter fields, different in different models, but in any case their
contribution to the purely gravitational sector of the action boils
down to a modification of the form factors of the basis invariants
with the Ricci tensor only. There are only two such in $S(2)$ and
ten in $S(3)$. Below, when referring to the purely gravitational
form factors, we shall assume that this reduction has already been
made. The full set of invariants for gravity and matter is considered
here because it is important that the maintenance of asymptotic
flatness be proved for the one-loop action in full generality. Apart
 from this, our main concern in the discussion below is the vacuum
action for the metric.

The functions $\gamma_i $ and $\Gamma_i $ in (2.7) and (2.8) are
the second-order and third-order form factors. The principal
assumption about these and higher-order form factors made in
the axiomatic approach is their analyticity which allows one to put
them in the spectral forms [4]. For example, the spectral form used
in [6] for the lowest-order form factors is

\begin{equation}
\gamma (-\Box) =  (\Box + \mu^2)^n \int\limits_0^{\infty}
 \frac{dm^2}{m^2 - \Box} \frac{w(m^2)}{(m^2 + \mu^2)^n} +
  \sum_{k=0}^{n-1} \frac{(-1)^k}{k!} (\Box + \mu^2)^k
{\left( \frac{\partial}{\partial \mu^2}\right)}^k
\gamma(\mu^2)
\end{equation}
where $w(m^2)$ is the spectral weight
\begin{equation}
w(m^2) = \frac{1}{2\pi i} \Bigl[ \gamma(-m^2 - i0)
                                                   - \gamma(-m^2 + i0) \Bigr] ,
\end{equation}
$\mu^2 > 0$ is an arbitrary parameter on which $\gamma(-\Box)$
does not depend, and $n$ is the degree of growth of
 $\gamma(-\Box)$ at large $\Box$ which will
presumably be fixed or bounded by the requirement of regularity
of the solution (see [1-3]). The requirement of asymptotic flatness
of the solution imposes restrictions only on the small-$\Box$
behaviours of the form factors [6,7]. In the small-$\Box$ limit,
eq. (2.10) reduces to the simple spectral form since the terms
modifying this form for $n > 0$ vanish in this limit [6]. On the
 other hand, this form should be generalized to allow for the
behaviour $1/\Box, |\Box| \to 0$ of $\gamma(-\Box).$
It remained unnoticed in paper [6] that the derivation in
this paper for the two purely gravitational second-order
form factors brings in fact to the following general result:
\begin{eqnarray}
\gamma_1(-\Box)& =& - \frac{2a}{\Box} - w_1(0) \log (-\Box)
+ O(1) ,\quad - \Box \to 0 \\
 \gamma_2(-\Box)& =&  \frac{a}{\Box} - w_2(0) \log (-\Box)
+ O(1) ,\quad - \Box \to 0
\end{eqnarray}
in which there appears an arbitrary constant $a$
\footnote{Eq. (39) of ref. [6] admits one more solution:
$ \Box \Bigl[ \gamma_1 (-\Box) + 3 \gamma_2(-\Box) \Bigr] \to a,
- \Box \to 0 $ which had been overlooked.}.
Only  the combination
\begin{equation}
\gamma_1(-\Box) + 2\gamma_2(-\Box) =
- \Bigl(w_1(0) + 2w_2(0) \Bigr) \log(-\Box) + O(1), \; -\Box \to 0
\end{equation}
should behave like $\log (-\Box)$ by the analysis in [6]. As will be
seen below, a similar situation takes place for the higher-order form
factors.

In the form factors $\gamma_i$ calculated from field theory, $a=0$
[9,15] since, by dimension, the terms $1/\Box$ cannot appear in the
loop expansion of the second-order form factors. They appear,
however, already in the third-order form factors $\Gamma_i$
[15]. For the generalized spectral forms of these form factors
see [10,12].

With the form factors in the spectral forms, the only nonlocal
structure that remains in $ S_{vac}$ is the inverse operator
$1/{(m^2 - \Box)}.$ This simplifies the procedure of obtaining
the expectation-value equations. When the action $S_{vac}$
is varied, the inverse operators are regarded as obeying the
variational rule
\begin{equation}
\delta \frac{1}{m^2 - \Box} = \frac{1}{m^2 - \Box} \delta\Box
\frac{1}{m^2 - \Box} \;  ,
\end{equation}
and, after the variation has been completed, all inverse operators
are replaced by the retarded Green functions
\footnote{For the derivation of this procedure in QFT see [8] and
references therein. In the phenomenological approach this set of
rules is taken for granted [4].}.
If
\begin{equation}
{\left.\frac{\delta S_{vac}}{\delta g_{\mu\nu}(x)}\right|}_{\Box
 \to \Box_{ret}}
\end{equation}
is to denote the result of this procedure, and a further notation is
introduced to separate the classical term of $S_{vac}:$
\begin{equation}
T^{\mu\nu}_{vac} \equiv \frac{2}{g^{1/2}}{\left. \frac{\delta
 S_{vac}}{\delta g_{\mu\nu}}\right|}_{\Box \to \Box_{ret}} -
\frac{2}{g^{1/2}} \frac{\delta S(1)}{\delta g_{\mu\nu}}\; ,
\end{equation}
then the expectation-value equations corresponding to the action
(2.1) are of the form
\begin{equation}
       R^{\mu\nu} - \frac{1}{2} g^{\mu\nu} R =
    8\pi \Bigl( T^{\mu\nu}_{vac} + T^{\mu\nu}_{source}\Bigr)
\end{equation}
where
\begin{equation}
        T^{\mu\nu}_{source} = \frac{2}{g^{1/2}}
\frac{\delta S_{source}}{\delta g_{\mu\nu}}  \;  ,
\end{equation}
and $T^{\mu\nu}_{vac}$ in (2.17)
can be interpreted as the energy-momentum
tensor of the in-vacuum. These equations are to be solved with zero
 initial data for the gravitational field at ${\cal I}^-.$  One arrives
at a Cauchy problem [2,3] for nonlocal equations with the retarded
kernels which are to be integrated from ${\cal I}^-$ to the future
until the solution hits a singularity if there remains one. One hopes
that it doesn't.

We do not consider the more general initial data, with a gravitational
wave at ${\cal I}^-,$ because in this case the action should also be
calculated more generally. Specifically, the solution of eq. (2.3)
can no more be taken in the form (2.4).
     \setcounter{equation}{0}
\section{The structure of $T^{\mu\nu}_{vac}$ at ${\cal I}^+$}
$\mbox{}$

It is natural to begin the study of the expectation-value
equations with the behaviour of $T^{\mu\nu}_{vac}$ at null
infinity since the nonlocal terms of the equations should be
responsible for the effect of the vacuum radiation.

At the future null infinity one has the Bondi-Sachs equation
[20,21] (see also appendix A) which is an exact consequence of
the expectation-value equations :
\begin{eqnarray}
- \frac{dM(u)}{du} = \frac{1}{4\pi} \int d^2 {\cal S}
\left[ \left(\frac{\partial}{\partial u} C_1 \right)^2 +
 \left(\frac{\partial}{\partial u} C_2 \right)^2 \right] +
\int d^2 {\cal S} \Bigl( \frac{1}{4} r^2 T^{\mu\nu}_{source}
\nabla_{\mu}v \nabla_{\nu}v + \nonumber \\
{} + \frac{1}{4} r^2 T^{\mu\nu}_{vac}
\nabla_{\mu}v \nabla_{\nu}v \Bigr) \biggl|_{{\cal I}^+} ,\qquad
\end{eqnarray}
\begin{eqnarray}
(\nabla u)^2 = 0, {(\nabla u, \nabla r)\Bigr|}_{{\cal I}^+} = -1\:, \\
(\nabla v)^2 = 0, {(\nabla u, \nabla v)\Bigr|}_{{\cal I}^+} = -2\:.
\end{eqnarray}
Here $u$ is the retarded time along ${\cal I}^+$ with the
natural normalization in (3.2), the integrals are over the
2-sphere ${\cal S}$  (normalized to have the area $4\pi$)
at which the null congruence $u=$const. crosses ${\cal I}^+,
r $ is the
luminosity distance along the rays of this congruence,
$M(u)$ is the Bondi mass, and ${\partial C_1}/{\partial u} ,
{\partial C_2}/{\partial u} $ are the Bondi-Sachs news
functions of the gravitational waves.

Eq. (3.1) is the conservation law missing in the theory of quantum
fields on a fixed gravitational background. In the collapse
problem, this is the backreaction equation relating "the
changing mass of the black hole" with the energy of the quanta
radiated by this black hole. In full quantum theory, both the
"black hole" and the quantum fields "on its background"
evolve from one and the same initial state, and one is able
to answer the question where does the black-hole radiation
take its energy from.
It takes it ultimately from the energy of the collapsing source
$T^{\mu\nu}_{source}$ which equals the ADM mass of the
expectation value of the metric and serves as an initial
datum $M(-\infty)$ for eq. (3.1).

The last term of eq. (3.1) is the flux of the vacuum energy
through ${\cal I}^+.$ For it to be finite, the flux component
of $T^{\mu\nu}_{vac}$ should decrease at ${\cal I}^+$
like $1/r^2.$ This is a necessary condition of asymptotic
flatness. Below, terms $O(1/r^3)$ in $T^{\mu\nu}_{vac}$
will be referred to as vanishing at ${\cal I}^+.$

When computing $T^{\mu\nu}_{vac}(x)$ at ${\cal I}^+$
from the action (2.5), all terms in which the Ricci curvature
appears at the observation point $x$ can be discarded
because $R_{\mu\nu}$ decreases at least like $1/r^2$
and will always be multiplied by a decreasing function.
Thus, the covariant derivatives $\nabla$ which appear
 in the basis invariants $\Re\Re\Re (i),$ etc. need not be
varied because the contributions of their variations to
$T^{\mu\nu}_{vac}$ contain the curvature at the
 observation point. For the same reason, in the expression
\begin{eqnarray}
\delta R^{\gamma}_{\nu} = \frac{1}{2} g^{\gamma\mu}
\Bigl( \nabla_{\mu}\nabla^{\alpha} \delta g_{\nu\alpha}
+\nabla_{\nu}\nabla^{\alpha} \delta g_{\mu\alpha} -
\nabla_{\mu}\nabla_{\nu} g^{\alpha\beta}
\delta g_{\alpha\beta} - \Box \delta g_{\mu\nu} + \nonumber \\
{} + 2 R_{\mu .\:.\:\nu}^{\;\:\alpha\beta} \delta g_{\alpha\beta}
\Bigr) +
 \frac{1}{2} R^{\gamma\alpha} \delta g_{\nu\alpha} -
\frac{1}{2} g^{\gamma\alpha} R^{\beta}_{\nu}
\delta g_{\alpha\beta} \; ,
\end{eqnarray}
the terms with the Ricci tensor can be discarded but the
term with the Riemann tensor cannot since the Riemann
tensor has components decreasing like $1/r.$ In (3.4), the
expression for $\delta R^{\gamma}_{\nu}$ has been brought
by commutations to the form used below.

Only the variations of the Ricci tensors and the variations of the
form factors in $S_{vac}$ can give nonvanishing contributions
to $T^{\mu\nu}_{vac}$ at ${\cal I}^+.$ It is easy to see that
the variation of a form factor
\begin{equation}
    \Gamma ( -\Box_1, -\Box_2,  ...  , -\Box_N )
\end{equation}
in the argument $\Box_p$ can contribute at ${\cal I}^+$ only
if, in this argument, the form factor behaves like $1/\Box_p ,
\Box_p \to -0.$ Thus, assuming $a=0$ in eqs. (2.12), (2.13), as
is the case in the field theoretic form factors, one can calculate
the variations of $\gamma_i(-\Box)$ in the action (2.7) by using
 the spectral form (2.10) in which, moreover, the terms
 appearing at $n > 0 $ can be disregarded when the
observation point tends to ${\cal I}^+.$ One finds
\begin{equation}
\int dx g^{1/2} \Re_1 \delta \gamma (-\Box) \Re_2 =
\int dx g^{1/2} \int\limits^{\infty}_{0} dm^2 w(m^2)
\Bigl(\frac{1}{m^2 - \Box} \Re_1 \Bigr) \delta \Box
\Bigl(\frac{1}{m^2 - \Box} \Re_2 \Bigr) \: .
\end{equation}
Since, by the result in [6],
\begin{eqnarray}
 \Bigl(\frac{1}{(m^2 - \Box_{ret})}\: \Re (x)\Bigr)\biggl|_{x
\to {\cal I}^+}  \propto
r^{-1}(x) \exp \Bigl( - |\mbox{const.}| m \sqrt{r(x)}\Bigr) (1 +
{\cal O}) , \\
{\cal O} \to 0, r(x) \to \infty, x \to {\cal I}^+,\nonumber
\end{eqnarray}
and
\begin{equation}
w(0) = \mbox{finite},
\end{equation}
one concludes that the contribution of $\delta \gamma (-\Box)$
to $T^{\mu\nu}_{vac}$ at ${\cal I}^+$ is $O(1/r^3).$ This
result applies to all form factor of the form (3.5) provided
that their behaviours in individual arguments are
$ O(\log(-\Box)), \Box \to -0.$ On the other hand, with the
behaviour
\begin{equation}
\Gamma (-\Box_1, -\Box_2, -\Box_3, ... ) =
\frac{1}{\Box_1} F(\Box_2, \Box_3, ... ) (1 + {\cal O}),
\quad {\cal O} \to 0, \Box_1 \to -0
\end{equation}
the variation of $\Box_1$ in the term
\begin{equation}
\int dx g^{1/2} \Gamma (-\Box_1, -\Box_2, -\Box_3, ... )
\Re_1 \Re_2 \Re_3 \;...
\end{equation}
of the action is asymptotically of the form
\begin{equation}
- \int dx g^{1/2} \Bigl(\frac{1}{\Box} \Re_1 \Bigr) \delta \Box
\frac{1}{\Box} \Bigl( F( \Box_2, \Box_3,  ... )\Re_2 \Re_3 \; .\:.\:.
\Bigr)
\end{equation}
and gives a nonvanishing contribution to $T^{\mu\nu}_{vac}$
at ${\cal I}^+$ proportional to $1/r^2.$

The fact that, in four dimensions, the variations of the
second-order form factors do not contribute to the energy
flux at infinity makes an important distinction of this case
from the case in two dimensions where $\gamma (-\Box)
\propto 1/\Box $ and the relevant contribution comes from
$\delta \gamma (-\Box)$ [3]. The variations of the third-order
 form factors $\Gamma_i $ in (2.8) can already contribute to
$T^{\mu\nu}_{vac}$ at ${\cal I}^+$ owing to the
$1/\Box, \Box \to -0 $ terms discovered in [15] but, to second
order in the curvature (in the equations), only the Ricci
tensors in the action $S(3)$ need to be varied.

For all terms of the vacuum action, in which the form factors
are of the  form (3.5)\footnote{ We do not
consider here the more general form
factors
$ \Gamma (-\Box_1, ... , -\Box_N, -\Box_{1+2},
-\Box_{1+3}, ... ) $
in which the operator arguments $\Box_{n+m}$ act on
products of two curvatures since such form factors appear
in $S_{vac}$ only beginning with $N=4$ [4,5].},
it is useful to introduce a quantity, {\it the generalized
current}, defined by varying the action with respect to the
Ricci tensor only. The sum of such terms in $S_{vac}$ can
be represented in the form
\begin{equation}
{\tilde S}_{vac} = \frac{1}{2(4\pi)^2} \sum_{N}
 \sum_{i} S_{i}(N) \;,
 \end{equation}
 \begin{equation}
S_{i}(N) =  \int dx g^{1/2} P_1(\nabla_1) .\:.\:. P_N(\nabla_N)
\Gamma (-\Box_1, ... , -\Box_N) R^{\cdot}_{\cdot}(x_1)
.\:.\:. R^{\cdot}_{\cdot}(x_N)
\end{equation}
where (3.13) is a contribution of the $N$th order basis invariant
number   " $i$ ", and the sum in (3.12) extends over both $i$ and
$N.$ In (3.13),$ R^{\cdot}_{\cdot} $ are the Ricci tensors with mixed
indices, and the polynomials in covariant derivatives $P_{n}
(\nabla_n)$ which are generally present in the tensor basis
[15,5] act on the respective $R^{\cdot}_{\cdot}(x_n)$ {\it after} the
action of the operator arguments $\Box_n$ of $\Gamma.$
In this
representation all operators $\Box_n$ in $\Gamma$ are
uniformly defined as applied to a mixed second-rank tensor,
and the advantage of taking it mixed is in the absence of
$g^{\mu\nu}$ factors contracting indices in (3.13) which
otherwise would need to be varied. These factors are
 generally present in the polynomials $P(\nabla)$ but,
in any case, their variations ( as well as the variation of
$g^{1/2}$ in the measure) do not contribute to $T^{\mu
\nu}_{vac}$ at ${\cal I}^+.$ The order of operations in (3.13)
is different from the one in (2.8) (cf. the explicit expressions in
[15]) but, at every order in the curvature, the action can be
brought to the form (3.13) by commutations.

Since the argument $\Box_n$ of $\Gamma$ is the first operator
acting on $R^{\cdot}_{\cdot}(x_n)$ in (3.13), it will be the last in the
variational derivative of (3.13) with respect
to $R^{\cdot}_{\cdot}(x_n).$ The
generalized current $I^{\mu}_{\nu}(\xi, x)$ is then defined by the
relation
\begin{equation}
\delta_{R}{\tilde S}_{vac} =
  \frac{1}{2(4\pi)^2} \int dx g^{1/2}
I^{\mu}_{\nu}(-\acton{17mm}{\Box, x) \delta R^{\nu}_{\mu}(x)} =
\frac{1}{2(4\pi)^2} \int dx g^{1/2}
I^{\mu}_{\nu}(-\acton{6mm}{\Box\:\strut, x\;}) \delta R^{\nu}_{\mu}(x)
\end{equation}
where the notation $\delta_{R}$ points out that only the Ricci
tensors in (3.13) are varied, and the argument $\Box$ of
$I^{\mu}_{\nu}(-\Box, x)$ is the argument of the form factor
$\Gamma$ in (3.13) that acts on the varied Ricci tensor. The
$I^{\mu}_{\nu}(\xi, x)$ is a tensor function of the spacetime
point $x$ and a function of a parameter $\xi$ which in eq.
(3.14) gets replaced by the operator $-\Box.$ This operator next
acts in either of the two ways pointed out in (3.14).

Given the action of the form (3.12), it is easy to calculate
$I^{\mu}_{\nu}(\xi, x)$ to each given order in the curvature.
For the action (2.5) we have
\begin{equation}
I^{\alpha}_{\beta}(\xi, x) = I^{\;\alpha}_{2\beta}(\xi, x) +
I^{\;\alpha}_{3\beta}(\xi, x) + O[\Re^3]
\end{equation}
where $I^{\;\alpha}_{2\beta}(\xi, x)$ and
$ I^{\;\alpha}_{3\beta}(\xi, x)$
are the contributions of $S(2)$ and $S(3)$ respectively, and
\begin{equation}
I^{\;\alpha}_{2\beta}(\xi, x) =  2 \gamma_1(\xi)
R^{\alpha}_{\beta}(x) + 2 \gamma_2 (\xi) \delta^{\alpha}_{\beta}
 R(x) \; ,
 \end{equation}
 \begin{equation}
 \begin{array}{ll}
I^{\;\alpha}_{3\beta}(\xi, x) = &
\;\;\;3 \Bigl[ \delta^{\mu}_{\nu} \Gamma_{10} ( \xi, -\Box_1, -\Box_2)
R^{\alpha}_{\mu}(x_1) R^{\nu}_{\beta}(x_2) \Bigr]_{x_1=x_2=x}
+  \\
   & {} + \Bigl[ \nabla_{1}^{\;\alpha}\nabla_{2\beta} \Gamma_{22}
( \xi, -\Box_1, -\Box_2) R(x_1) R(x_2)\Bigr]_{x_1=x_2=x } - \\
 & {} - 2\delta^{\alpha}_{\beta} \nabla_{\mu} \Bigl[ \nabla_{2}^{\;\nu}
\Gamma_{22}( -\Box_1, -\Box_2, \xi ) R^{\mu}_{\nu}(x_1)
R(x_2) \Bigr]_{x_1=x_2=x} + .\:.\:.
\end{array}
\end{equation}
where only the purely gravitational terms are written down. We
do not present the latter expression in full but exemplify it with
the contributions of two third-order invariants, number 10 and
number 22 (see [15]). Eq. (3.17) illustrates the general structure
of the current $I^{\mu}_{\nu}(\xi, x).$

A remarkable property of $I^{\mu}_{\nu}(\xi, x)$ is that the
variations of both the Ricci tensors and the form factors in (3.12)
are expressed entirely through this quantity. Indeed, in addition
to (3.14), we have
\begin{equation}
\delta_{\Gamma}{\tilde S}_{vac} = \frac{1}{2(4\pi)^2} \int
dx g^{1/2} \delta \Box' \frac{1}{\Box'- \Box''}
\Bigl[ I^{\nu}_{\mu}(-\Box', x'') - I^{\nu}_{\mu}(-\Box'', x'') \Bigr]
R^{\mu}_{\nu}(x')
\end{equation}
where the notation $\delta_{\Gamma}$ points out that only the
form factors $\Gamma$ in (3.13) are varied, and we have used
the general formula for a variation of an operator function
[12,14]:
\begin{equation}
\int dx g^{1/2} A \left(\delta f(\Box)\right) B =
\int dx g^{1/2} \delta \Box_{B} \frac{f(\Box_A) -
 f(\Box_B)}{\Box_A - \Box_B} A B \; .
\end{equation}
It is understood that $\Box_A$ (or $\Box_B$) is the operator
$\Box$ acting on $A$ (or $B$), and similarly in (3.18) $\Box'$
acts on $x',$ and $\Box''$ on $x''$ with subsequently setting
$x'=x''=x.$ The operators $\delta \Box_B$ and $\Box_B,$
and similarly $\delta \Box'$ and $\Box'$ in (3.18), do not
commute and act in the indicated order. The identity
\begin{equation}
\int dx g^{1/2} ( \delta \Box'' - \delta \Box' ) F(x',x'') =
\int dx g^{1/2} (-\delta \log g^{1/2}) ( \Box'' - \Box' )
F(x',x'')
\end{equation}
(with an arbitrary two-point tensor $F(x', x'')$ contracting
into  a scalar at $x'= x''$) serves to check that varying the
left-hand side of the equality
\begin{equation}
\int dx g^{1/2} f(\Box_B) A B = \int dx g^{1/2} f(\Box_A)
A B
\end{equation}
with the aid of eq. (3.19) gives the same result as varying its
right-hand side.

Although the generalized current $I^{\mu}_{\nu}(\xi, x),$
as calculated from the action (2.5), is given in the form of an
expansion, it enters the expectation-value equations as a single
whole and determines the vacuum stress at null infinity.
Indeed, since (3.14) and (3.18) are the only contributions
surviving in (2.16) when the observation point tends to
${\cal I}^+, T^{\mu\nu}_{vac}$ at ${\cal I}^+$ is obtained
in a closed form. As seen from (3.16), (3.17), the behaviour
of $I^{\mu}_{\nu}(\xi, x)$ in $\xi$ includes all the behaviours
of the form factors in individual arguments. The
 $I^{\mu}_{\nu}(\xi, x)$ may, therefore, be of a significance
in axiomatic theory. For this current as a function of its
 parameter  argument one must postulate the existence of a
spectral form similar to (2.10) but allowing for the
spectral weight to have a $\delta(m^2) $ singularity at
$m^2 = 0.$ Assuming for simplicity $n = 0$ in (2.10)
(the modification concerning the large $\Box$ is irrelevant
to the present discussion) we set
\begin{equation}
I^{\mu}_{\nu}(\xi, x) = \int\limits^{\infty}_{-\epsilon}
\frac{dm^2}{m^2 + \xi}\: w^{\mu}_{\nu}(m^2, x) \; ,
\end{equation}
and then eqs. (3.14) and (3.18) take the form
\begin{eqnarray}
\delta_{R}{\tilde S}_{vac} = \frac{1}{2(4\pi)^2}
\int dx g^{1/2} \int\limits^{\infty}_{-\epsilon} dm^2
\delta R^{\nu}_{\mu}(x)\Bigl[ \frac{1}{m^2 - \Box}\:
 w^{\mu}_{\nu}(m^2, x)\Bigr] \; , \\
\delta_{\Gamma}{\tilde S}_{vac} = \frac{1}{2(4\pi)^2}
\int dx g^{1/2} \int\limits^{\infty}_{-\epsilon} dm^2
\Bigl[\delta \Box \frac{1}{m^2 - \Box}
R^{\nu}_{\mu}(x) \Bigr] \Bigl[ \frac{1}{m^2 - \Box}
 \:w^{\mu}_{\nu}(m^2, x)\Bigr] \; .
\end{eqnarray}
Along with expression (3.4) and an easily derivable
expression for $\delta \Box $ in (3.24) they determine
$T^{\mu\nu}_{vac}$ at ${\cal I}^+.$
    \setcounter{equation}{0}
   \section{The asymptotic flatness constraints }
  $\mbox{}$

    Even with the $1/\xi, \xi \to 0$ behaviour of $I^{\mu}_{\nu}(\xi, x),$
the contribution to $T^{\mu\nu}_{vac}$ coming from (3.24) is
$O(1/r^2)$ at ${\cal I}^+.$ Also the contributions coming from
the term with the Riemann tensor and the term with the $\Box$
operator in (3.4) are $O(1/r^2).$ We have, therefore, from (3.14)
and (3.4)
\begin{eqnarray}
T^{\mu\nu}_{vac}(x) = \frac{1}{2(4\pi)^2} \Bigl(
\nabla^{\mu}\nabla_{\alpha}I^{\alpha\nu}
(-\acton{9mm}{\Box_{ret},\strut x\;}) +
\nabla^{\nu}\nabla_{\alpha}I^{\alpha\mu}
(-\acton{9mm}{\Box_{ret},\strut x\;}) - \qquad\qquad \\
{} - g^{\mu\nu} \nabla_{\beta}\nabla_{\alpha}I^{\alpha\beta}
(-\acton{9mm}{\Box_{ret},\strut x\;}) \Bigr) +
\; O(r^{-2}(x)),\;\; x \to {\cal I}^{+} \quad  \nonumber
\end{eqnarray}
\begin{equation}
I^{\alpha\beta}(\xi, x) = g^{(\alpha \gamma}(x) I^{\beta)}_{
\gamma}(\xi, x)\; .
\end{equation}
The terms of order $r^{-2}(x)$ in (4.1) are the ones to be retained
but it is not our purpose here to calculate $T^{\mu\nu}_{vac}$
\footnote{
For this calculation to lowest order in the curvature see [6,7]. The
cancellation of the $1/\Box$ terms being established in the present
paper, this calculation can now be done to second order in the
 curvature on the basis of the results in [15].}.
The point is that the remaining terms in (4.1) should also behave
like $O(r^{-2}(x))$ for the asymptotic flatness to be maintained.
This behaviour should , moreover, hold at every order in the
curvature because even small disturbances of the metric can
violate the asymptotic flatness. This does not mean, however,
that the function
\begin{equation}
I^{\alpha\mu}(-\acton{9mm}{\Box_{ret},\strut x\;})
\end{equation}
should behave like $O(r^{-2}(x));$ it suffices that
\begin{equation}
\nabla_{\alpha}I^{\alpha\mu}(-\acton{9mm}{\Box_{ret},\strut x\;})
 = O(r^{-2}(x))\;, \;\; x \to {\cal I}^+.
\end{equation}
Owing to this fact, the $1/\xi, \xi \to 0$ behaviour of $I^{\alpha
\mu}(\xi, x)$ is not completely ruled out but condition (4.4) imposes
a constraint on the coefficient of this behaviour.

For implementing this constraint the derivative $\nabla_{\alpha}$
in (4.4) should be commuted with the operator $\Box_{ret}:$
\begin{eqnarray}
\nabla_{\alpha}I^{\alpha\mu}(-\acton{9mm}{\Box_{ret},\strut x\;}) =
\int\limits^{\infty}_{-\epsilon} dm^2
\Bigl( \frac{1}{m^2 - \Box_{ret}} \nabla_{\alpha}
w^{\alpha\mu}(m^2, x) + \qquad\quad \\
\qquad\qquad{} + \frac{1}{m^2 - \Box_{ret}}
[\nabla_{\alpha}, \Box] \frac{1}{m^2 - \Box_{ret}}
w^{\alpha\mu}(m^2, x) \Bigr)\; . \nonumber
\end{eqnarray}
Here all operators act to the right on $x,$ and
\begin{equation}
w^{\mu\nu}(m^2, x) = g^{(\mu\gamma}(x)\;
w^{\nu)}_{\gamma}(m^2, x)\; .
\end{equation}
Both terms in (4.5) are generally $O(r^{-1}(x))$ but the
commutator term contains an extra power of the curvature.
If we denote
\begin{equation}
w^{\mu}(m^2, x) \equiv  \nabla_{\alpha}w^{\alpha\mu}(m^2, x)
+ [\nabla_{\alpha}, \Box]\: \frac{1}{m^2 - \Box_{ret}}\:
w^{\alpha\mu}(m^2, x)
\end{equation}
and
\begin{equation}
I^{\mu}(\xi, x) \equiv  \int\limits^{\infty}_{0}
\frac{dm^2}{m^2 + \xi}\: w^{\mu}(m^2, x)\; ,
\end{equation}
so that
\begin{equation}
\nabla_{\alpha}I^{\alpha\mu}
(-\acton{9mm}{\Box_{ret},\strut x\;}) =
I^{\mu}(-\acton{9mm}{\Box_{ret},\strut x\;})\; ,
\end{equation}
then, by (4.4) and the result in [6], the behaviour of the
function (4.8) should already be
\begin{equation}
I^{\mu}(\xi, x) = -\; w^{\mu}(0, x) \log \xi + O(1)\;, \; \xi \to 0
\end{equation}
\begin{equation}
w^{\mu}(0, x) = \mbox{finite.}
\end{equation}

The vector current (4.8) is obtained from (3.15) as an expansion:
\begin{equation}
I^{\mu}(\xi, x) = I_2^{\mu}(\xi, x) + I^{\mu}_3(\xi, x) +
O[\Re^3]
\end{equation}
where $I^{\mu}_2$ and $I^{\mu}_3$ are the contributions
of the actions $S(2)$ and $S(3).$ For the contribution of the
action $S(2)$ one finds by using eq. (3.16):
\begin{eqnarray}
I^{\mu}_2(-\acton{9mm}{\Box_{ret},\strut x\;})
= \Bigl(\gamma_1(-\Box_{ret}) +
2\gamma_2(-\Box_{ret})\Bigr)\nabla^{\mu}R +
2\:[\nabla_{\alpha}, \gamma_1(-\Box_{ret})] R^{\alpha\mu} +
\quad \\
 {} + 2\:[\nabla^{\mu}, \gamma_2(-\Box_{ret})] R\; , \nonumber
\end{eqnarray}
and
\begin{equation}
I^{\mu}_2(\xi, x) = \Bigl( \gamma_1(\xi) + 2\gamma_2(\xi)
\Bigr) \nabla^{\mu}R(x) + O[\Re^2]
\end{equation}
whence the constraint (2.14) for the second-order form factors
immediately follows. Should the constant $a$ in (2.12),(2.13)
be nonvanishing, the commutator terms in (4.13) would contribute
to the constraint condition for the third-order form factors.
Since, however, $a=0$ as discussed above, these commutator
 terms are $O(1/r^2)$ as seen from their spectral forms.

The constraint condition for the third-order form factors is thus
of the form
\begin{equation}
I_3^{\mu}(-\acton{9mm}{\Box_{ret},\strut x\;})
+ O[\Re^3] = O\Bigl( r^{-2}(x)\Bigr)
\; ,\;\; x \to {\cal I}^+
\end{equation}
and
\begin{equation}
I_3^{\mu}(\xi, x) = \nabla_{\alpha}I_3^{\alpha\mu}(\xi, x) +
O[\Re^3]
\end{equation}
since, in this case, the commutator of $\nabla_{\alpha}$ with
$\xi = -\Box$ contributes to $O[\Re^3]$ already. On the other
 hand, with the field-theoretic form factors [15] one obtains
\begin{eqnarray}
I_3^{\alpha\mu}(\xi, x)& =& \frac{1}{\xi} A^{\alpha\mu}(x) +
(\log \xi) B^{\alpha\mu}(x) + O(\xi^0)\;, \\
I_3^{\mu}(\xi, x)& =& \frac{1}{\xi} A^{\mu}(x) + O(\log \xi) \; ,
\xi \to 0\; ,\\
A^{\mu}(x)& =& \nabla_{\alpha}A^{\alpha\mu}(x) + O[\Re^3]
\end{eqnarray}
where the coefficients $A^{\mu}(x),$ etc. can be expanded
in some vector basis second-order in the curvature:
\begin{equation}
A^{\mu}(x) = \tr \sum_{p}A_p(\Box_1, \Box_2)
{\Re_1\Re_2}^{\: \mu}(p) + O[\Re^3]\;.
\end{equation}
Examples of the basis structures in (4.20) are
\begin{eqnarray}
{\Re_1\Re_2}^{\:\mu}(1)& =& \nabla^{\mu}R_1\cdot R_2 {\hat 1}\;, \\
{\Re_1\Re_2}^{\:\mu}(2)& =& \nabla^{\mu}R_1^{\alpha
\beta}\cdot\nabla_{\alpha}\nabla_{\beta}R_2 {\hat 1}\;, \\
{\Re_1\Re_2}^{\:\mu}(3)& =& {\hat {\cal R}}_1^{\alpha\beta}\nabla^{\mu}
{\hat {\cal R}}_{2\alpha\beta} \;, \\
{\Re_1\Re_2}^{\:\mu}(4)& =&  R_1^{\mu\alpha}
\nabla_{\alpha}{\hat P}_2 \;,
\end{eqnarray}
etc. where all curvatures of the set (2.2) participate, and
the trace in (4.20) refers to the matrices ${\hat 1},{\hat P},$
etc. (cf. a similar construction of the basis of invariants in
[15]). We do not present the basis in (4.20) in full although
it is important to have it in full for obtaining the results
below. In order that (4.15) hold for any choice of the
in-state \footnote{
With the specifications made in sec. 2, this choice
boils down to the choice of $T^{\mu\nu}_{source}$
in eq. (2.18). Variations in $T^{\mu\nu}_{source}$
induce variations in the curvature of the solution.
Eq. (4.15) should, therefore, hold for any configuration
of the curvature. It is also important for inferring (4.25) that,
 by the construction of the curvature basis in the action [5],
$I_3^{\alpha\mu}(\xi, x)$ can have no total derivative
terms of the form $\Box X^{\alpha\mu}(\xi, x)$ or
$\nabla^{(\alpha}X^{\mu)}(\xi, x).$ Hence $A^{\alpha\mu}(x)$
can have no such terms.},
there should be
\begin{equation}
A^{\mu}(x) + O[\Re^3] = 0
\end{equation}
and hence
\begin{equation}
A_p(\Box_1, \Box_2) = 0\; .
\end{equation}

Eqs. (4.26) are the constraints to be satisfied by the coefficients
of the $1/\Box$ asymptotic behaviours of the third-order
form factors. Let us introduce a notation for these
coefficients:
\begin{eqnarray}
\Gamma_{i}(-\Box_1,-\Box_2, -\Box_3) =
\frac{1}{\Box_1}F^1_i(\Box_2, \Box_3) + O(\log(-\Box_1))\;,
\Box_1 \to -0 \\
\Gamma_{i}(-\Box_1,-\Box_2, -\Box_3) =
\frac{1}{\Box_2}F^2_i(\Box_3, \Box_1) + O(\log(-\Box_2))\;,
\Box_2 \to -0 \\
\Gamma_{i}(-\Box_1,-\Box_2, -\Box_3) =
\frac{1}{\Box_3}F^3_i(\Box_1, \Box_2) + O(\log(-\Box_3))\;,
\Box_3 \to -0
\end{eqnarray}
where there appear functions of two variables $F^m_i$ with
$m=1$ to 3 and $i=1$ to 29. The functions $A_p$ in (4.20)
are certain linear combinations of the $F^m_i.$ By an
explicit calculation with the action $S(3)$ in (2.8) and [15]
one can work up these combinations to see if they vanish.
The commutator and potential curvatures appear in the
basis in (4.20) but the contributions of $\delta
{\hat {\cal R}}_{\mu\nu}$ and $\delta {\hat P}$ to (3.14) may,
in this
calculation, be omitted since all $F^m_i$ with $\Box_m$
acting on ${\hat {\cal R}}_{\mu\nu} $ or ${\hat P}$ vanish
[12,15]. This makes it possible to carry out the check of
asymptotic flatness for a generic quantum field model.

The results are as follows. Of $3\times 29$ functions $F^m_i$
only 21 in the table of ref. [15] do not vanish and are not
related to each other by the symmetries of the form factors.
With the 21 nonvanishing $F^m_i$ the expansion (4.20)
gives rise to 14 constraints (4.26) which, by linearly combining
them, can be brought to the following form:
\begin{eqnarray}
F^3_{25}(\Box_1, \Box_2)& =& \frac{1}{2}(\Box_1 - \Box_2)
F^3_{28}(\Box_1, \Box_2) \; ,\\
F^1_{10}(\Box_1, \Box_2)& =& -\frac{1}{12}(\Box_2 - \Box_1)^2
F^3_{28}(\Box_1, \Box_2) \; , \\
F^1_{28}(\Box_1, \Box_2)& =& 2 F^3_{27}(\Box_1, \Box_2)  +
\frac{3}{2}(\Box_2 - \Box_1)F^1_{29}(\Box_1, \Box_2) \; ,\\
F^3_{24}(\Box_1, \Box_2)& =&\frac{1}{2}F^1_{25}(\Box_1, \Box_2)
- \frac{1}{4}(\Box_1 - \Box_2)F^1_{28}(\Box_2, \Box_1) \; , \\
F^3_{22}(\Box_1, \Box_2)& =& - \frac{1}{2}F^3_{24}(\Box_1, \Box_2)
+ \frac{1}{8}(\Box_1 + \Box_2)F^1_{28}(\Box_1, \Box_2) \; , \\
F^1_{11}(\Box_1, \Box_2)& =& - \frac{1}{4}(\Box_2 - \Box_1)
F^1_{23}(\Box_1, \Box_2)
\;, \\
F^1_{23}(\Box_1, \Box_2)& =& -2 F^3_{22}(\Box_1, \Box_2) +
(\Box_1 - \Box_2)F^1_{27}(\Box_1, \Box_2) + \\
 & {} +{} & \frac{1}{4}(\Box_1 + \Box_2)F^1_{28}(\Box_2, \Box_1) +
 \frac{3}{4}(\Box_2 - \Box_1)\Box_1 F^1_{29}(\Box_1, \Box_2)
\; , \nonumber\\
F^1_9(\Box_1, \Box_2)& =& \frac{1}{32}(\Box_1 + \Box_2)
(\Box_1 - \Box_2)^2 F^1_{29}(\Box_1, \Box_2) - \\
  & & {} - \quad\frac{1}{24}
(\Box_1^2 + \Box_2^2) F^3_{27}(\Box_1, \Box_2) \; ,\nonumber \\
F^1_{22}(\Box_1, \Box_2)& =& - \frac{1}{4}F^1_{25}(\Box_1, \Box_2)
+ \frac{3}{8}(\Box_1 - \Box_2)^2 F^1_{29}(\Box_1, \Box_2) \; ,\\
F^1_5(\Box_1, \Box_2)& =& \frac{1}{4}(\Box_1 - \Box_2)^2
F^1_{26}(\Box_1, \Box_2) \; , \\
F^1_{16}(\Box_1, \Box_2)& =& \quad(\Box_1 - \Box_2)
F^1_{26}(\Box_1, \Box_2) \; , \\
F^1_8(\Box_1, \Box_2)& =& \frac{1}{4}(\Box_1 + \Box_2)
F^1_{21}(\Box_1, \Box_2) \; , \\
F^1_{18}(\Box_1, \Box_2)& =& \frac{1}{2}
F^1_{21}(\Box_1, \Box_2) \; , \\
F^1_{19}(\Box_1, \Box_2)& =& - \frac{1}{4}
F^1_{21}(\Box_1, \Box_2) \; .
\end{eqnarray}
It is now a matter of a direct inspection to check if the
$F^m_i$ calculated in [15] satisfy these constraints.
They do!

Relations (4.30) - (4.43) leave only  7 independent
nonvanishing $F^m_i$ for which one can take the
functions $ F^1_{21}, F^1_{25}, F^1_{26}, F^1_{27},
 F^3_{27}, F^3_{28},
F^1_{29}.$ With the exception of $F^1_{26}$ and $F^1_{27},$ these
functions are symmetric in their $\Box$ arguments: $F^1_{25},
F^3_{27}$ and $F^3_{28}$ are symmetric owing to the respective
symmetries of the form factors $\Gamma_{25}, \Gamma_{27}$
and $\Gamma_{28},$ and the symmetry of $F^1_{21}, F^1_{29}$
is a property of the explicit expressions in [15].

By expressing all $F^m_i$ through the 7 independent ones, one
can bring the coefficient $A^{\alpha\mu}(x)$ in (4.17) to the
form
\begin{equation}
A^{\mu\nu}(x) = - \nabla_{\alpha} \nabla_{\beta}
K^{[\mu\alpha][\nu\beta]}(x) + O[\Re^3]
\end{equation}
in which the fulfilment of condition (4.25) is manifest, and the
function $K^{\mu\alpha\nu\beta}(x)$ which appears
 antisymmetrized in (4.44) is of the following form:

\begin{eqnarray}
\; K^{\mu\alpha\nu\beta}(x)& =& 3 \tr F^1_{29}(\Box_1, \Box_2)
\Bigl\{ 4 \nabla^{\mu}\nabla^{\nu} R^{\gamma\sigma}_1
\cdot \nabla_{\gamma}\nabla_{\sigma}R^{\alpha\beta}_2 +
2 ( \Box_2 - \Box_1) \Bigl[ \nabla^{(\mu}R_1 \cdot
\nabla^{\nu)}R^{\alpha\beta}_2 \Bigr] - \nonumber\\
& & {} - \Box_2 (\Box_2 - \Box_1) g^{\mu\nu} \Bigl[ R_1^{\alpha\beta}
\cdot R_2 \Bigr] \Bigr\}{\hat 1} + {}
\tr F^3_{28}(\Box_1, \Box_2)\Bigl\{ 4\nabla^{\mu}R^{\alpha
\gamma}_1 \cdot \nabla^{\nu}R^{\:\:\beta}_{2\gamma}\Bigr\}
{\hat 1} + \nonumber \\
 & & {}+ \tr F^1_{27}(\Box_1, \Box_2)\Bigl\{ 8\nabla^{\mu}
\nabla^{\nu}R_1^{\alpha\beta} \cdot R_2 \Bigr\}
{\hat 1} +  8\;\tr F^1_{26}(\Box_1, \Box_2)\Bigl\{
 \nabla^{\mu}\nabla^{\nu}R_1^{\alpha\beta} \cdot
{\hat P}_2  \Bigr\} + \nonumber\\
& & {} + \tr F^1_{25}(\Box_1, \Box_2)\Bigl\{
4 R_1^{\mu\beta}\cdot R^{\nu\alpha}_2 + 2g^{\mu\nu}
R_1^{\alpha\beta}\cdot R_2 \Bigr\} {\hat 1} + \nonumber \\
& & {}+ \frac{1}{2}\tr F^1_{21}(\Box_1, \Box_2)\Bigl\{
{\hat {\cal R}}_1^{\mu\alpha}\cdot {\hat {\cal R}}_2^{\nu\beta}
\Bigr\} +{} \tr F^3_{27}(\Box_1, \Box_2)\Bigl\{
8 \nabla^{(\mu}R_1^{\nu)\gamma}\cdot \nabla_{\gamma}
R_2^{\alpha\beta} + \nonumber \\
& & {} + 4\:g^{\mu\nu}\nabla_{\gamma}R_1^{
\sigma\alpha}\cdot \nabla_{\sigma}R_2^{\gamma\beta}
+ 4 \nabla^{(\mu}R_1 \cdot \nabla^{\nu)}R_2^{\alpha
\beta} + \\
& & {} + 2 (\Box_1 - \Box_2) g^{\mu\nu}\Bigl[ R_1 \cdot R_2^{\alpha
\beta}\Bigr] + g^{\mu\nu} \nabla^{\alpha}\nabla^{\beta}R_1
\cdot R_2 \Bigr\}{\hat 1} \; .\nonumber
\end{eqnarray}
The one-loop expressions for the functions  $F$ entering
(4.45) are given in [15] but one may conjecture that the
7 independent structures in (4.45) and (4.44) (5 in the case
of pure gravity) is the general result  independent of models
and approximations (although this remains to be checked by
repeating the analysis above for the general form of the action).
As will be seen below, expression (4.45) gets considerably
simplified when inserted in the formula for the news functions.
    \setcounter{equation}{0}
    \section{ The news functions. Contribution of the}
    \section*{third-order form factors}
    $\mbox{} $

A significance of the $1/\Box$ asymptotic terms in the vacuum
form factors is that they contribute to the energy of the outgoing
gravitational waves. This energy is the term with the news
functions ${\partial C_1}/{\partial u}, {\partial C_2}/{\partial u}$
in the Bondi-Sachs equation (3.1).

Obtaining the news functions requires solving the dynamical
equations already. However, there is a short cut : eq. (2.4).
We may use the fact that the news functions appear as a
coefficient of the $1/r$ behaviour of the Riemann tensor
at null infinity. Indeed, we have (see appendix A)
\begin{equation}
\nabla_{\alpha}v \nabla_{\mu}v\: m_{\beta} m_{\nu}
R^{\alpha\beta\mu\nu}\biggl|_{{\cal I}^+} =
-\frac{8}{r}\: \frac{\partial^2}{{\partial u}^2}\:{\bf C}
+ O\left( \frac{1}{r^2}\right)
\end{equation}
where
\begin{equation}
{\bf C} = C_1 + i C_2 \; ,
\end{equation}
and $m_\beta$ is a complex null vector tangent to the
2-sphere ${\cal S}$:
\begin{equation}
(m, \nabla u) = (m, \nabla v) = (m, m) = 0 ,\quad (m, m^*) = -2
\end{equation}
with $ \nabla u, \nabla v$ and ${\cal S}$ in (3.1)-(3.3), and
$m^*$ complex conjugate to $m.$ The contribution of the
outgoing gravitational waves to the mass loss, eq. (3.1), is
then
\begin{equation}
\frac{1}{4\pi} \int d^2 {\cal S} \left|\: \frac{\partial}{
\partial u}\: {\bf C}\:\right|^2 \; ,
\end{equation}
and
\begin{equation}
-\frac{\partial^2}{{\partial u}^2}\: {\bf C}
= \left\{\frac{r}{8}\:\nabla_{\alpha}v \nabla_{\mu}v \:
m_{\beta} m_{\nu}R^{\alpha\beta\mu\nu}\right\}
\Biggl|_{{\cal I}^+}\;.
\end{equation}

On the other hand, the Riemann tensor can be calculated
with the aid of eq. (2.4). Its $1/r$ behaviour is then
obtained as {\it the leading asymptotic behaviour of the retarded
Green function}. Using (2.4b) we have
\begin{equation}
-\frac{\partial^2}{{\partial u}^2}\: {\bf C}
= \left\{\frac{r}{2}\nabla_{\alpha}v \nabla_{\mu}v \:
m_{\beta} m_{\nu} \nabla^{[\mu}\nabla^{<\alpha}
\frac{1}{\Box}\Bigl( R^{\nu]\beta>} + O[R^2_{..}]
\Bigr)\right\}\biggl|_{{\cal I}^+}
\end{equation}
where the quadratic terms are to be copied from (2.4b),
and the quadratic terms with no overall $1/\Box$
factor - all terms in the curly brackets in (2.4b) - do not contribute.

Expression (5.6) can next be simplified as follows. The derivatives
in (5.6) appear projected either as $\nabla^{\nu}v \nabla_{\nu}$
or as $m^{\nu} \nabla_{\nu}.$ In both cases the projected
derivatives can be commuted with the remaining factors of
$\nabla v$ and $m$ since the components of $\nabla \nabla
v$ and $\nabla m$ in the null tetrad basis are $O(1/r)$ at
${\cal I}^+.$ The projected derivatives become then
acting on some scalar $X$ which behaves like $1/r$ at
${\cal I}^+,$ and in this case
\begin{eqnarray}
m^{\nu} \nabla_{\nu} X\biggl|_{{\cal I}^+} & =
 &  O\left( \frac{1}{r^2} \right) \; , \\
\nabla^{\nu}v \nabla_{\nu}X\biggl|_{{\cal I}^+} &
= & -2\: \frac{\partial}{\partial u}\: X + O\left(
\frac{1}{r^2} \right)
\end{eqnarray}
(see appendix A). In this way one obtains
\begin{equation}
\frac{\partial^2}{{\partial u}^2}\: {\bf C}
= -\: \frac{\partial}{\partial u}\:
\left\{\:\frac{\partial}{\partial u}\; \frac{r}{2} \:
m_{\beta} m_{\nu}\: \frac{1}{\Box}\:\Bigl( R^{\nu\beta}
 + O[R^2_{..}] \Bigr)\right\}\Biggl|_{{\cal I}^+} \; .
\end{equation}

Eq. (5.9) can now be integrated over $u$ from $-\infty$ to
a given point of ${\cal I}^+$ to obtain the news functions.
Since, at $u = -\infty, {\partial {\bf C}}/{\partial u} = 0,$
there remains to be shown that the expression in the curly
brackets in (5.9) also vanishes at $u = -\infty.$ This is
shown in appendix B. As a result one obtains
\begin{eqnarray}
\frac{\partial}{\partial u}\: {\bf C}& = &
- \frac{\partial}{\partial u}\: \Bigl\{\: \frac{r}{2}\:m_{\beta}
m_{\nu} \frac{1}{\Box}\Bigl[ R^{\nu\beta} +
(\nabla^{\nu} \frac{1}{\Box}R^{\gamma\delta})(\nabla^{\beta}
\frac{1}{\Box}R_{\gamma\delta}) - \qquad\\
 &  &\qquad\qquad\quad\qquad {} - 2(\nabla_{\gamma} \frac{1}{\Box}R^{\nu\delta}
)(\nabla_{\delta} \frac{1}{\Box}R^{\beta\gamma})
+ O[R^3_{..}] \Bigr] \Bigr\} \Biggl|_{{\cal I}^+}\nonumber
\end{eqnarray}
where for the Ricci tensor one can use the dynamical
equations \footnote{ This method can also be applied to the
classical problems in the gravitational radiation.}.

There is, of course, a classical gravitational radiation or,
more generally, a radiation induced by the classical source
$T^{\mu\nu}_{source}$ in eq. (2.18). The respective
contribution\footnote{ Even this contribution is not
purely classical since the metric to be used is the
solution of the
expectation-value equations. The same concerns the flux
 of $T^{\mu\nu}_{source}$ at ${\cal I}^+$ in eq. (3.1).
It would be interesting to consider a case where $T^{\mu
\nu}_{source}$ does not induce the gravitational waves
but $T^{\mu\nu}_{vac}$ does.}
to the news functions (call it ${\partial {\bf C}_{source
}}/{\partial u}$) is obtained by substituting for $R^{\nu\beta}$
its Einstein value
\begin{equation}
R^{\nu\beta}_{cl} = 8\pi \left(  T^{\nu\beta}_{source} -\:
\frac{1}{2}\: g^{\nu\beta} T_{source} \right)
\end{equation}
in all terms of (5.10) including $ O[R^3_{..}].$
The remaining contributions in (5.10) stand for the gravitational
radiation induced by the vacuum stress,
\begin{equation}
\frac{\partial}{\partial u}\: {\bf C} =
\frac{\partial}{\partial u}\: {\bf C}_{source} + \:
\frac{\partial}{\partial u}\: {\bf C}_{vac}\; .
\end{equation}
By (5.10) and the dynamical equations (2.18)\footnote{
Since, at ${\cal I}^-,$ the flux components of $T^{\mu\nu}_{vac}$
vanish, eq. (5.13) is valid even without the limitation implied in
(2.4b) (see sec. 2).},
\begin{equation}
\frac{\partial}{\partial u}\: {\bf C}_{vac} =
- \frac{\partial}{\partial u}\: \Bigl\{ 4\pi r\: m_{\beta} m_{\nu}
\frac{1}{\Box}\Bigl(\: T^{\nu\beta}_{vac} +\:
O[R^2_{..}] \Bigr) \Bigr\}\biggl|_{{\cal I}^+}
\end{equation}
and
\begin{equation}
T^{\mu\nu}_{vac} = T^{\mu\nu}_{vac}(2) +
 T^{\mu\nu}_{vac}(3) + O[R^3_{..}]
\end{equation}
where $T^{\mu\nu}_{vac}(2)$ and $T^{\mu\nu}_{vac}(3)$
are the contributions of the actions $S(2)$ and $S(3)$ in (2.5).

In (5.10) and (5.13) there appears an expression of the form
\begin{equation}
\Bigl\{ r\:m_{\beta} m_{\nu} \frac{1}{\Box} X^{\nu\beta}
\Bigr\} \biggl|_{{\cal I}^+}
\end{equation}
with some tensor $X^{\mu\nu},$ and we have used the fact
that a contribution of this form with
\begin{equation}
X^{\mu\nu} = g^{\mu\nu} X
\end{equation}
vanishes by the orthogonality relation in (5.3). Another property
of expression (5.13) which will be used below is that a contribution
of the form (5.15) with
\begin{equation}
X^{\mu\nu} = \nabla^{\mu} X^{\nu} +
\nabla^{\nu} X^{\mu}
\end{equation}
is of a higher order in the curvature :
\begin{equation}
O[ X^{\mu}\times R_{..} ] \; .
\end{equation}
Indeed, in this case,
\begin{equation}
m_{\beta}m_{\nu}\: \frac{1}{\Box} X^{\nu\beta} =
2 m_{\beta}m_{\nu} \nabla^{\nu} \frac{1}{\Box} X^{\beta} +
2 m_{\beta}m_{\nu} \frac{1}{\Box} \Bigl(
[\nabla^{\nu}, \Box] \frac{1}{\Box} X^{\beta}\Bigr) \; .
\end{equation}
By (5.7), the first term of this expression is $O(1/r^2),$
and the remaining term contains a commutator.

To lowest order in the curvature, only the
contribution of the action $S(2)$ is to be considered. By (3.4)
and (3.14),
\begin{eqnarray}
T^{\mu\nu}_{vac}(2)  & =  & \frac{1}{2(4\pi)^2}
\Bigl( \nabla^{\mu}\nabla_{\alpha}I^{\alpha\nu}_2 +
\nabla^{\nu}\nabla_{\alpha}I^{\alpha\mu}_2  -
g^{\mu\nu}\nabla_{\beta}\nabla_{\alpha}I^{\alpha\beta}_2
 - \qquad   \\
  &  & \qquad\qquad\qquad\qquad\qquad\qquad\qquad\quad
  {}  - \Box I^{\mu\nu}_2 \Bigr) + O[ R^2_{..} ]\nonumber
\end{eqnarray}
where
\begin{equation}
I^{\mu\nu}_{2} = I_{2}^{\mu\nu}(-
\acton{9mm}{\Box_{ret},\strut x\;}) \; ,
\end{equation}
and the explicit form of $I_2(\xi, x)$ is given in (3.16). The
first two terms in (5.20) are of the form (5.17), and, therefore,
their contribution to (5.13) is $O[R^2_{..}].$ The third term in
(5.20) is of the form (5.16), and, therefore, its contribution to
(5.13) vanishes. In the remaining term of (5.20),the
$\Box$ operator kills the Green function:
\begin{equation}
\frac{\partial}{\partial u}\: {\bf C}_{vac} =
\:\frac{\partial}{\partial u}\: \Bigl\{\: \frac{r}{8\pi}\:
m_{\beta}m_{\nu}I_2^{\nu\beta}(
-\acton{9mm}{\Box_{ret},\strut  x\;})
\Bigr\}\biggl|_{{\cal I}^+} + O[R^2_{..}] \; ,
\end{equation}
{\it and this is the reason why at all the asymptotic behaviours
of the form factors at small} $\Box$ {\it are relevant to the
gravitational waves.} Since the second-order form factors
$\gamma_1, \gamma_2$ behave like $\log (-\Box)$ at small
$\Box,$ we have
\begin{equation}
I_2^{\nu\beta}( -
\acton{9mm}{\Box_{ret},\strut x\;}) = O\left( \frac{1}{r^2}\right)
\;,\qquad x \to {\cal I}^+
\end{equation}
and the contribution (5.22) vanishes. Thus the vacuum
contribution to the news functions begins with second
order in the curvature:
\begin{equation}
\frac{\partial}{\partial u}\: {\bf C}_{vac} = O[R^2_{..}]\; .
\end{equation}

At second order in the curvature both $S(2)$ and $S(3)$
contribute to ${\bf C}_{vac}.$ We are presently interested
in the contribution of $S(3)$ which we shall denote
${\bf C}_{vac}(3).$ Since the action $S(3)$ is cubic in the
 curvature, the accuracy in (5.13) is sufficient for calculating
this contribution :
\begin{equation}
\frac{\partial}{\partial u}\: {\bf C}_{vac}(3) =
\:-\: \frac{\partial}{\partial u}\: \Bigl\{ 4\pi r\: m_{\beta} m_{\nu}
\frac{1}{\Box}\Bigl( T^{\nu\beta}_{vac}(3) +
O[R^3_{..}] \Bigr) \Bigr\}\biggl|_{{\cal I}^+}~~~.
\end{equation}
To second order in the curvature, $T^{\mu\nu}_{vac}(3)$
is of the form similar to (5.20) :
\begin{eqnarray}
T^{\mu\nu}_{vac}(3)  & =  & \frac{1}{2(4\pi)^2}
\Bigl( \nabla^{\mu}\nabla_{\alpha}I^{\alpha\nu}_3 +
\nabla^{\nu}\nabla_{\alpha}I^{\alpha\mu}_3  -
g^{\mu\nu}\nabla_{\beta}\nabla_{\alpha}I^{\alpha\beta}_3
 - \qquad   \\
  &  &\qquad\qquad\qquad\qquad\qquad\qquad\qquad\quad
  {} - \Box I^{\mu\nu}_3 \Bigr) + O[ R^3_{..} ]  \nonumber
\end{eqnarray}
with
\begin{equation}
I^{\mu\nu}_3 = I^{\mu\nu}_3(-\acton{9mm}{\Box_{ret},\strut x\;})\; .
\end{equation}
Again, the first two terms in (5.26) are of the form (5.17), and their
contribution to (5.25) is $O[R^3_{..}].$ Again the third term is
proportional to the metric, and its contribution vanishes.
Again the $\Box$ operator in the remaining term kills the
Green function:
\begin{equation}
\frac{\partial}{\partial u}\: {\bf C}_{vac}(3) =
\:\frac{\partial}{\partial u}\: \Bigl\{\: \frac{r}{8\pi}\:
m_{\beta}m_{\nu}I_3^{\nu\beta}(-
\acton{9mm}{\Box_{ret},\strut x\;})
\Bigr\}\biggl|_{{\cal I}^+} +\; O[R^3_{..}]
\end{equation}
but this time the contribution (5.28) does not vanish
because the third-order form factors behave like $1/\Box$
at small $\Box.$ By (4.17),
\begin{equation}
I^{\nu\beta}_{3}(-\acton{5mm}{\Box,\strut x\;}) =
- \Bigl(\acton{14mm}{\frac{1}{\Box}\Bigr) A^{\nu\beta}(x)}
 + \log(-\acton{13mm}{\Box)\strut B^{\nu\beta}(x)}
 +\;  O\left(\frac{1}{r^3}\right)\; ,
\end{equation}
and only the term with $1/\Box$ survives in (5.28). With the
expression  for $A^{\nu\beta}(x)$ given in (4.44) the result is
\begin{equation}
\frac{\partial}{\partial u}\: {\bf C}_{vac}(3) =
\:\frac{\partial}{\partial u}\: \Bigl\{\: \frac{r}{8\pi}
\:m_{\mu}m_{\nu}\:\frac{1}{\Box} \nabla_{\alpha}
\nabla_{\beta}K^{[\mu\alpha][\nu\beta]}(x) +
O[R^3_{..}] \Bigr\}\biggl|_{{\cal I}^+} \; .
\end{equation}

Expression (4.45) for $K^{\mu\alpha\nu\beta}$ can now
be simplified by using that (i) all terms containing the metric
with the indices of $K^{\mu\alpha\nu\beta}$ can be
discarded since any contraction among $m_{\mu}m_{\nu}
\nabla_{\alpha} \nabla_{\beta}$  in (5.30) results in a vanishing
contribution, and (ii) any derivative $\nabla$ with the
indices of $K^{\mu\alpha\nu\beta}$ can be treated
as in integration by parts because the respective total
derivative contracts with either $m$ or $\nabla,$ and
its contribution vanishes. As a result, $K^{\mu\alpha
\nu\beta}$ in (5.30) can be replaced by the following
expression :
\begin{eqnarray}
& &{\tilde K}^{\mu\alpha\nu\beta}(x) =  12 \tr F^1_{29}(
\Box_1, \Box_2) \Bigl[\nabla^{\mu}\nabla^{\nu}R^{\gamma
\sigma}_1 \cdot \nabla_{\gamma}\nabla_{\sigma} R^{\alpha
\beta}_2 \Bigr] {\hat 1} +  \\
 & & {} + 4 \tr F^3_{28}(\Box_1, \Box_2)
 \Bigl[\nabla^{\mu}R^{\alpha\gamma}_1 \cdot\nabla^{\nu}
 R^{\;\beta}_{2\gamma} \Bigr] {\hat 1} +
    8 \tr F^3_{27}(\Box_1, \Box_2)
 \Bigl[\nabla^{(\mu}R^{\nu)\gamma}_1
 \cdot \nabla_{\gamma}R^{\alpha\beta}_2 \Bigr] {\hat 1} +
 \nonumber \\
   & & {} + 4 \tr F^1_{25}(\Box_1, \Box_2) \Bigl[ R_1^{\mu\beta}\cdot
R_2^{\nu\alpha} \Bigr] {\hat 1}
   +  \tr \Bigl\{ 6(\Box_2 - \Box_1) F^1_{29}(\Box_1, \Box_2) +
   8 F^1_{27} (\Box_1, \Box_2) -  \nonumber \\
    & & {} - 4 F^3_{27}(\Box_1, \Box_2)\Bigr\}
\Bigl[ \nabla^{\mu}\nabla^{\nu} R^{\alpha\beta}_1 \cdot R_2
\Bigr] {\hat 1} + \;
   8 \tr F^1_{26}(\Box_1, \Box_2) \Bigl[ \nabla^{\mu}
\nabla^{\nu} R^{\alpha\beta}_1 \cdot {\hat P}_2 \Bigr] +\nonumber \\
& & \qquad\qquad\qquad {} + \frac{1}{2} \tr F^1_{21}(\Box_1, \Box_2)
\Bigl[ {\hat {\cal R}}^{\mu\alpha}_1 \cdot
{\hat {\cal R}}^{\nu\beta}_2 \Bigr]\; .\qquad\qquad\nonumber
\end{eqnarray}
    \setcounter{equation}{0}
\section{ The vacuum news functions in the lowest}
\section*{nonvanishing approximation}
$\mbox{} $

To complete the calculation of the vacuum news functions
in the lowest nonvanishing approximation we must consider
the contribution of the action $S(2):$
\begin{eqnarray}
\frac{\partial}{\partial u}\: {\bf C}_{vac}(2)& =&
- \frac{\partial}{\partial u}\Bigl\{ 4 \pi r m_{\beta}
m_{\gamma} \frac{1}{\Box} \Bigl[ T^{\beta\gamma}_{vac}(2)\;
+ 2 \Bigl( \nabla^{\beta} \frac{1}{\Box}R^{cl}_{\mu\nu}\Bigr)
\Bigl(\nabla^{\gamma}\frac{1}{\Box} T^{\mu\nu}_{vac}(2)\Bigr) -
\nonumber\\
& & {} - \Bigl( \nabla^{\beta} \frac{1}{\Box}R^{cl}\Bigr)
\Bigl(\nabla^{\gamma}\frac{1}{\Box} T_{vac}(2)\Bigr) -
\: 4 \Bigl( \nabla_{\sigma} \frac{1}{\Box}R_{cl}^{\beta\mu}\Bigr)
\Bigl(\nabla_{\mu}\frac{1}{\Box} T^{\gamma\sigma}_{vac}(2)\Bigr) +
\nonumber \\
& & {} + 2 \Bigl( \nabla^{\gamma} \frac{1}{\Box}R_{cl}^{\beta\mu}\Bigr)
\Bigl(\nabla_{\mu}\frac{1}{\Box} T_{vac}(2)\Bigr)\Bigr] +
O( \hbar^2 ) + O[ R^3_{..} ] \Bigr\}\biggl|_{{\cal I}^+}\; .
\end{eqnarray}
Here $R^{\mu\nu}_{cl}$ is the notation in (5.11), and in the nonlinear
terms of eq. (5.10) we omitted the contributions $T_{vac}\times
T_{vac} = O(\hbar^2)$ of second order in the Planck constant. As
shown above, expression (6.1) is of second order in the curvature;
the contribution of first order in this expression vanishes.

The expression (5.20) for $T^{\mu\nu}_{vac}(2)$ completed
with terms $O[ R^2_{..} ]$ is of the form
\begin{eqnarray}
T^{\mu\nu}_{vac}(2)& = & \frac{1}{2(4\pi)^2} \Bigl(
\nabla^{\mu}\nabla_{\alpha} I_2^{\alpha\nu} +
\nabla^{\nu}\nabla_{\alpha} I_2^{\alpha\mu} -
g^{\mu\nu} \nabla_{\beta}\nabla_{\alpha}I_2^{\alpha
\beta} - \Box I_2^{\mu\nu} +  \\
  & &\qquad\qquad\qquad{}  + 2 R_{\alpha .\:.\:\beta}^{\:\:\mu\nu}
  I_2^{\alpha\beta}
{} + \frac{1}{2}g^{\mu\nu} R_{\alpha\beta}
 I_2^{\alpha\beta}\Bigr) + \Pi^{\mu\nu} \nonumber
\end{eqnarray}
where
\begin{equation}
\Pi^{\mu\nu} = \frac{2}{g^{1/2}}\frac{\delta_{\gamma}S(2)}{
\delta g_{\mu\nu}}\biggl|_{\Box \to \Box_{ret}}
\end{equation}
with
\begin{equation}
\delta_{\gamma}S(2) = \frac{1}{2(4\pi)^2} \int dx g^{1/2}
\Bigl( R^\mu_\nu\: \delta \gamma_1( -\Box) R^\nu_\mu +
R\: \delta \gamma_2( -\Box) R \Bigr)
\end{equation}
is the contribution of the variations of the lowest-order form
factors (see eqs. (3.19), (3.24)). When inserting expression (6.2)
 in the linear term of (6.1), all terms in (6.2) proportional to
$g^{\mu\nu}$ or to the $\Box$ operator can be omitted, and
for the contribution of the first two terms in (6.2) one can use
eq. (5.19).

In the nonlinear terms of (6.1), the expression (6.2) is needed
only up to $O[ R^2_{..}].$ By using (3.16), we obtain
\begin{eqnarray}
T^{\mu\nu}_{vac}(2) = \frac{1}{(4\pi)^2} \Bigl[
\nabla^{\mu}\nabla^{\nu} \Bigl(\gamma_1(-\Box) +
2\gamma_2(-\Box)\Bigr) R - \Box \gamma_1(-\Box)
R^{\mu\nu} - \nonumber \\
{} - \frac{1}{2}\: g^{\mu\nu} \Box \Bigl( \gamma_1(-\Box) +
 4 \gamma_2(-\Box) \Bigr) R \Bigr] + O[ R^2_{..} ] \; .
\end{eqnarray}
This makes it possible to calculate also the Riemann tensor
with accuracy $O[R^2_{..}] :$
\begin{eqnarray}
R^{\alpha\beta\mu\nu} = R^{\alpha\beta\mu\nu}_{cl}
 - \frac{2}{\pi} \gamma_1(-\Box) \nabla^{[\mu}\nabla^{
<\alpha}R^{\nu] \beta>} + \frac{1}{4\pi}\Bigl(
\gamma_1(-\Box) + 2\gamma_2(-\Box)\Bigr) \times \nonumber \\
\times \Bigl( g^{\nu\beta} \nabla^{\mu}\nabla^{\alpha} -
 g^{\nu\alpha} \nabla^{\mu}\nabla^{\beta} -
g^{\mu\beta} \nabla^{\nu}\nabla^{\alpha} +
g^{\mu\alpha} \nabla^{\nu}\nabla^{\beta}\Bigr) R
+ O[ R^2_{..} ] \; .
\end{eqnarray}
Here we used eq. (2.4a) which defines also
\begin{equation}
R^{\alpha\beta\mu\nu}_{cl} = \frac{1}{\Box} \Bigl(
2 \nabla^{\mu}\nabla^{<\alpha}R^{\nu\beta>}_{cl} -
2 \nabla^{\nu}\nabla^{<\alpha}R^{\mu\beta>}_{cl} +
O[ R^2_{..} ] \Bigr) \; .
\end{equation}

Finally, by combining the results above, the following
 expression is obtained for the contribution
${\partial {\bf C}_{vac}(2)}/{\partial u}$ to the news
functions:
\begin{eqnarray}
\frac{\partial}{\partial u}\: {\bf C}_{vac}(2)& = &
- \frac{\partial}{\partial u} \biggl\{ \frac{r}{4\pi}
m_{\mu}m_{\nu} \frac{1}{\Box}\Bigl(
\gamma_1(-\Box_2) \Bigl[ 4\nabla_{\alpha}\frac{1}{\Box}
R^{\mu\beta}_{1\: cl}\cdot \nabla_{\beta}R^{\nu\alpha}_{
2\: cl} + \\
& & {} + 4 \nabla^{\alpha}\nabla^{\mu} \frac{1}{\Box} R^{\beta
\nu}_{1\: cl}\cdot R^{cl}_{2\alpha\beta} -
2\nabla^{\alpha}\nabla^{\beta}\frac{1}{\Box}R^{\mu\nu}_{
1\: cl}\cdot R^{cl}_{2\alpha\beta} +
R^{\mu\nu}_{1\: cl}\cdot R_{2\: cl} \Bigr] + \nonumber\\
& &{}  + \Bigl( \gamma_1(-\Box_2) + 2 \gamma_2(-\Box_2)
\Bigr)\Bigl[ \nabla^{\mu}\frac{1}{\Box}R_{1\: cl}\cdot
\nabla^{\nu}R_{2\: cl} - 2 \nabla^{\nu}\frac{1}{\Box}
R^{\mu\alpha}_{1\: cl}\cdot \nabla_{\alpha}R_{2\: cl}+ \nonumber \\
& &{} +  \frac{1}{2}\frac{1}{\Box}R^{\mu\nu}_{1\: cl}\cdot
\Box  R_{2\: cl}
{} - \frac{1}{2}R^{\mu\nu}_{1\: cl}\cdot R_{2\: cl} \Bigr] +
(4\pi)^2 \Pi^{\mu\nu}\Bigr) + O(\hbar^2) +
O[ R^3_{..} ] \biggr\} \Biggl|_{{\cal I}^+} \; .\nonumber
\end{eqnarray}
This contribution involves the lowest-order vacuum form
 factors in the whole range of their dependence on the
$\Box$ argument.

The total result is
\begin{equation}
\frac{\partial}{\partial u}\:{\bf C} = \frac{\partial}{\partial u}
\:{\bf C}_{source} +
\frac{\partial}{\partial u}\:{\bf C}_{vac}(2) +
 \frac{\partial}{\partial u}\:{\bf C}_{vac}(3) + O(\hbar^2) +
O[ R^3_{..} ]
\end{equation}
with ${\partial {\bf C}_{vac}(2)}/{\partial u}$ in (6.8) and
 ${\partial {\bf C}_{vac}(3)}/{\partial u}$ in (5.30). Note
that the news functions appear squared in the mass-loss
formula (3.1), and the vacuum contribution to
${\partial {\bf C}}/{\partial u}$ begins with second order
in the curvature, eq. (5.24). Therefore, in the absence of a
classical radiation, the energy of the vacuum gravitational
waves is of  order $O[ \hbar^2 R^4_{..} ]$ which makes
this effect difficult to be noticed in perturbation theory.
\begin{center}
\underline{\bf Acknowledgments}
\end{center}

The authors are grateful to B.S.DeWitt and L.P.Grishchuk
for helpful discussions.

The present work was supported in part by the Russian
Foundation for Fundamental Research (Grant 93-02-15594),
 the International (Soros) Science Foundation (Grant
MQY000), the EC grant INTAS-93-493, and the
NATO travel grant CRG 920991.
    \setcounter{equation}{0}
    \renewcommand{\theequation}{A.\arabic{equation}}
    \section*{Appendix A. The asymptotically flat metric
   at null infinity}
   $\mbox{} $

The general asymptotically flat metric in a chart covering
${\cal I}^+$ is built by considering a congruence of null
hypersurfaces $u =$ const. generated by the light rays
reaching ${\cal I}^+.$ The generators are labeled by
two parameters $\theta, \varphi$ taking values on a
2~-~sphere ${\cal S},$ and the luminosity distance $r$
is used as a parameter along the generators. The metric
is then of the form [21]
\begin{equation}
ds^2 = - V du^2 + 2 \Psi du dr + r^2 f_{ab}(dx^{a} -
U^a du)(dx^b - U^b du)
\end{equation}
where
\begin{eqnarray}
a, b = 1,2 ;& &\quad\quad     x^1 = \theta,\; x^2 = \varphi\; ,
\qquad\qquad \nonumber \\
f_{ab}dx^a dx^b& = &\frac{1}{2}(e^{2\gamma} + e^{2\delta})
d\theta^2 + (e^{\gamma - \delta} - e^{\delta - \gamma})
\sin \theta d\theta d\varphi +  \\
& &\qquad\qquad\quad \qquad\qquad\qquad\qquad {}
+\: \frac{1}{2}(e^{-2\gamma} + e^{-2\delta})\sin^2\theta
d\varphi^2   \nonumber
\end{eqnarray}
and
\begin{eqnarray}
& & \frac{1}{\Psi} = (\nabla u, \nabla r) < 0 \; , \\
& & \Psi \biggl|_{{\cal I}^+} = -1\; .
\end{eqnarray}
In this metric, $(\nabla u)^2 = 0$ is to ensure that the
hypersurfaces $u =$ const. are null, $(\nabla x^a, \nabla u
) = 0 $ is to ensure that the lines $u =$ const.,$x^a =$ const.
are null geodesics, det$ f_{ab} = \sin^2 \theta$ is to ensure
that $r$ is the luminosity parameter along these geodesics,
condition (A.3) is to ensure that this parameter is monotonic,
and condition (A.4) is to choose the retarded time $u$
 coincident with the proper time of an observer at large
and constant $r.$

At the limit of ${\cal I}^+ (r \to \infty, u =$ const., $
\theta = $const., $\varphi = $const. ) the metric behaves as
follows [21]:
\begin{eqnarray}
V &=&  1 - \frac{2{\cal M}}{r} + O\left(\frac{1}{r^2}\right)\; , \\
\frac{\gamma + \delta}{2}& =& \frac{C_1}{r} +
 O\left(\frac{1}{r^2}\right)\; , \\
\frac{\gamma - \delta}{2}& =& \frac{C_2}{r} +
 O\left(\frac{1}{r^2}\right)\; , \\
U^a& =& \frac{2N^a}{r^2} + O\left(\frac{1}{r^3}\right) \; ,\\
\Psi& =& -1 - \frac{2B}{r^2} + O\left(\frac{1}{r^3}\right)
\end{eqnarray}
where ${\cal M}, C_1, C_2, N^a, B$ are functions of $\theta, \varphi,
u.$

The $C_1, C_2 $ differentiated with respect to $u$ are the
Bondi-Sachs news functions [20, 21]. In the gauge (A.1)
they stand for the radiation degrees of freedom of the
gravitational field. The Bondi mass ${M}(u)$ is obtained
 by averaging the coefficient ${\cal M}$ in (A.5)
over the unit 2-sphere :
\begin{equation}
M(u) = \frac{1}{4\pi} \int d^2{\cal S} {\cal M} =
\frac{1}{4\pi} \int\limits_0^{2\pi}d\varphi \int\limits_0^{\pi}
d\theta \sin \theta {\cal M}(\theta, \varphi, u)\; .
\end{equation}
Its limiting value at $u \to - \infty$ is the ADM mass, and the
difference
\begin{equation}
M(-\infty) - M(u) = \int\limits_{-\infty}^{u} du \Bigl(
- \frac{dM}{du}\Bigr)
\end{equation}
is the energy radiated away through ${\cal I}^+$ by the
instant $u$ of retarded time ( see [22] and references
therein).

Some components of the Riemann and Ricci tensors calculated
in the metric (A.1) - (A.9) at ${\cal I}^+$ are as follows
\footnote{By our calculations, the equation $R_{rr} =
(2/r)(\Psi^{-1}\partial_{r}\Psi - (\partial_{r}\gamma)^2)$
( in the present notation) given in [21] for the axisymmetric
case is in error. We obtain $ R_{rr} =
(2/r)\Psi^{-1}\partial_{r}\Psi - 2(\partial_{r}\gamma)^2.$
It follows from the former equation that, for $R_{\mu\nu}$
of a compact spatial support, $\Psi = -1 + O(1/r^3).$ With our
result, $\Psi = -1 + O(1/r^2).$ The significance of the
behaviour of $\Psi$ is seen from eqs. (A.15), (A.18).}:
\begin{eqnarray}
R_{u\theta u\theta}\biggl|_{{\cal I}^+}& =&
{} - r\: \frac{\partial^2}{{\partial u}^2}\:C_1 + O(1)\; ,\\
R_{u\theta u\varphi}\biggl|_{{\cal I}^+}& = &
{} - r \sin \theta\: \frac{\partial^2}{{\partial u}^2}\:C_2 + O(1)\; ,\\
R_{u\varphi u\varphi}\biggl|_{{\cal I}^+}& =&
 r \sin^2 \theta\:\frac{\partial^2}{{\partial u}^2}\:C_1 + O(1)\; ,\\
R_{rr}\biggl|_{{\cal I}^+}& =&
{} -  \frac{2}{r^4}\Bigl(C_1^2 + C_2^2 + 4B\Bigr) +
 O\left(\frac{1}{r^5}\right)\; ,\\
 R_{r\theta}\biggl|_{{\cal I}^+}& = &
 \: - \:\frac{1}{r^2}\:\Bigl[ 2N^{\theta} +
 \:\frac{1}{\sin^2 \theta}\:\frac{\partial}{\partial \theta}
 \Bigl(C_1 \sin^2 \theta\Bigr) + \frac{\partial}{\partial
 \varphi}\:\Bigl(\frac{C_2}{\sin \theta}\Bigr) \Bigr]
 {} + O\Bigl(\frac{1}{r^3}\Bigr) \; , \\
R_{r\varphi}\biggl|_{{\cal I}^+}& = &
 \: - \:\frac{1}{r^2}\:\Bigl[ 2N^{\varphi}\sin^2 \theta +
 \:\frac{1}{\sin \theta}\:\frac{\partial}{\partial \theta}
 \Bigl(C_2 \sin^2 \theta\Bigr) - \frac{\partial}{\partial
 \varphi}\: C_1 \Bigr] + O\Bigl(\frac{1}{r^3}\Bigr) \; , \\
 R_{ur}\biggl|_{{\cal I}^+}& =&
 {} \:\frac{1}{r^3}\:\frac{\partial}{\partial u}\:
 \Bigl(C_1^2 + C_2^2 + 4B\Bigr) + O\Bigl(\frac{1}{r^4}\Bigr)\;,\\
 R_{u\theta}\biggl|_{{\cal I}^+}& = &
 \:  \:\frac{1}{r}\:\frac{\partial}{\partial u}\:\Bigl[ 2N^{\theta} +
 \:\frac{1}{\sin^2 \theta}\:\frac{\partial}{\partial \theta}
 \Bigl(C_1 \sin^2 \theta\Bigr) + \frac{\partial}{\partial
 \varphi}\:\Bigl(\frac{C_2}{\sin \theta}\Bigr) \Bigr]
 {} + O\Bigl(\frac{1}{r^2}\Bigr) \; , \\
R_{u\varphi}\biggl|_{{\cal I}^+}& = &
 \:  \:\frac{1}{r}\:\frac{\partial}{\partial u}\:
 \Bigl[ 2N^{\varphi}\sin^2 \theta +
 \:\frac{1}{\sin \theta}\:\frac{\partial}{\partial \theta}
 \Bigl(C_2 \sin^2 \theta\Bigr) - \frac{\partial}{\partial
 \varphi}\: C_1 \Bigr] + O\Bigl(\frac{1}{r^2}\Bigr) \; , \\
 R_{uu}\biggl|_{{\cal I}^+}& =&
 {} - \frac{2}{r^2}\:\Bigl[\frac{\partial}{\partial u}\:
 {\cal M} + \Bigl( \frac{\partial}{\partial u}\:C_1\Bigr)^2 +
 {} \Bigl( \frac{\partial}{\partial u}\:C_2\Bigr)^2 +
 {} \frac{1}{\sin \theta}\:\partial_{a} \Bigl(
 \sin \theta\:\frac{\partial}{\partial u}\: N^{a}
 \Bigr) \Bigr] + \nonumber \\
& & \qquad\qquad\quad\qquad\qquad\qquad\qquad
{} + O\Bigl(\frac{1}{r^3}\Bigr) \; .
\end{eqnarray}

When referring to tensors at ${\cal I}^+$ we always mean their
projections on the null tetrad $\nabla u, \nabla v, m, m^{*}$
introduced in (3.2), (3.3) and (5.3) where $v$ and $m_{\alpha}$
are asymptotically of the form
\begin{eqnarray}
v\Bigl|_{{\cal I}^+}& =& 2r + u + O\left(\frac{1}{r}\right)\; ,\\
m_{\alpha}\Bigl|_{{\cal I}^+}& =& r(\nabla_{\alpha}\theta +
i\sin \theta \nabla_{\alpha}\varphi) + O(1)\; .
\end{eqnarray}
The null-tetrad components of physical quantities are regular
at ${\cal I}^+$ i.e. are either finite or decreasing like inverse
powers of $r.$ This is true specifically of tensors obtained by
the action of the retarded form factors (see below). The
null-tetrad vectors may be regarded as covariantly constant
at ${\cal I}^+$ since the null-tetrad components of their
derivatives are $O(1/r).$ Thus, up to curvature terms,
\begin{equation}
\nabla_{\mu}\nabla_{\alpha}v\biggl|_{{\cal I}^+} =
\frac{1}{2r} (m_{\mu}m_{\alpha}^{*} +
 m^{*}_{\mu}m_{\alpha}) + O[R_{..}]\;,
\end{equation}
and the curvature terms are $O(1/r).$ Eqs. (5.7), (5.8) are
obtained by calculating the derivatives projected on the
null tetrad in terms of the Bondi-Sachs coordinates.
Specifically, eq. (5.7) owes to the fact that
$ m^{\alpha}\Bigl|_{{\cal I}^+} = O(1/r)$ as seen from
(A.23) and (A.1).

It follows from the asymptotic expressions above that the
null-tetrad components of the curvature tensor decrease
at ${\cal I}^+$ like $1/r$ or faster. Specifically,
\begin{equation}
\nabla_{\alpha}v \nabla_{\mu}v\: m_{\beta}m_{\nu}
R^{\alpha\beta\mu\nu}\biggl|_{{\cal I}^+} =
-\: \frac{8}{r}\:\Bigl(\frac{\partial^2}{{\partial u}^2}C_1 +
i\frac{\partial^2}{{\partial u}^2}C_2 \Bigr) +
O\left(\frac{1}{r^2}\right)
\end{equation}
which is eq. (5.1). The energy flux component of the Ricci
 tensor at ${\cal I}^+$
\begin{equation}
R_{\mu\nu}\nabla^{\mu}v \nabla^{\nu}v\biggl|_{
{\cal I}^+} = 4R_{uu}\biggl|_{{\cal I}^+}
\end{equation}
is given in eq. (A.21). By averaging (A.21) over the unit
2-sphere one obtains the relation
\begin{equation}
- \frac{dM(u)}{du} =     \frac{1}{4\pi} \int d^2{\cal S} \Bigl[
\Bigl( \frac{\partial}{\partial u}C_1 \Bigr)^2 +
\Bigl( \frac{\partial}{\partial u}C_2 \Bigr)^2 \Bigr] +
\frac{1}{8\pi} \int d^2{\cal S}\:\frac{1}{4}\: r^2
 R^{\mu\nu}\nabla_{\mu}v \nabla_{\nu}v\biggl|_{{\cal I}^+}
\end{equation}
which, after using the dynamical equations
\begin{equation}
R^{\mu\nu} - \frac{1}{2}g^{\mu\nu}R = 8\pi T^{\mu\nu}_{total}
\; ,
\end{equation}
becomes the conservation law (3.1). Here $T^{\mu\nu}_{total}$
is the total energy-momentum tensor which in eq. (2.18) is
\begin{equation}
T^{\mu\nu}_{total} = T^{\mu\nu}_{source} +
T^{\mu\nu}_{vac}\; .
\end{equation}
    \setcounter{equation}{0}
   \renewcommand{\theequation}{B.\arabic{equation}}
   \section*{Appendix B. The retarded Green function in
   the past of ${\cal I}^+$}
   \setcounter{equation}{0}
$\mbox{}$

To lowest order in the curvature, the retarded operator $1/\Box$
acting on an arbitrary tensor source $X^{\alpha_1 \ldots \alpha_k}$
is of the following form:
\begin{eqnarray}
\lefteqn{
- \frac{1}{\Box}X^{\alpha_1 \ldots \alpha_k}(x) = }  &   \\
    &=  \frac{1}{4\pi} \int\limits_{{past\: of}\; x}
  d{\bar x} {\bar g}^{1/2}
\delta ( \sigma( x, {\bar x} )) g^{\alpha_1}_{\;\;{\bar \alpha}_1}
( x, {\bar x}) \ldots g^{\alpha_k}_{\;\;{\bar \alpha}_k}( x,
{\bar x}) X^{{\bar \alpha}_1 \ldots {\bar \alpha}_k}({\bar x})
+ O[X \times \Re] \nonumber
\end{eqnarray}
where $\sigma(x, {\bar x})$ is the world function [23], or
geodetic interval biscalar [24], $ g^{\alpha}_{\;\;{\bar
 \alpha}}( x, {\bar x}) $ is the geodetic parallel displacement
bivector [24], and the integration point ${\bar x}$ is in the
past of the observation point $x.$ Here and below, the bar
over a symbol means that this symbol refers to the point
${\bar x}.$

It follows from a comparison of the equations defining
$ g^{\alpha}_{\;\;{\bar \alpha}}( x, {\bar x}) $ with
the ones defining $\sigma(x, {\bar x})$ that, up to the
curvature terms, the parallel displacement bivector can
be calculated as follows:
\begin{equation}
g^{\alpha}_{\;\;{\bar \alpha}}( x, {\bar x})  =
- \nabla^{\alpha} {\bar \nabla}_{\bar \alpha}
\sigma (x, {\bar x}) + O[R_{..}]
\end{equation}
whence it also follows that
\begin{equation}
\nabla_{\mu}
g^{\alpha}_{\;\;{\bar \alpha}}( x, {\bar x})  =
 O[R_{..}]\; .
\end{equation}
If $\ell^{\mu}_{i}$ with $i=1$ to 4 are the vectors of the
null tetrad, then the null-tetrad components of the tensor
(B.1) are obtained by calculating the contractions
\begin{equation}
\ell^{\mu}_{i}(x) g_{\mu{\bar \mu}}( x, {\bar x}) \; .
\end{equation}
By using eq. (B.2) and a perturbative expression for the
world function, it is easy to see that, when $x$ tends to
${\cal I}^+,$ and ${\bar x}$ is in a compact domain,
the contractions (B.4) remain finite. Hence the null-tetrad
components of the tensor (B.1) decrease at ${\cal I}^+$
like $O(1/r)$ - the fact assumed in the main text. The
 expression for the massive retarded Green function
is similar to (B.1) [6]. Therefore, generally, the null-tetrad
components of tensors obtained by the action of the
retarded form factors are regular at ${\cal I}^+.$

As seen from (B.1), we are always dealing with some scalar
source
\begin{equation}
{\cal Y}({\bar x}) =
g^{\alpha_1}_{\;\;{\bar \alpha}_1}
( x, {\bar x}) \ldots g^{\alpha_k}_{\;\;{\bar \alpha}_k}( x,
{\bar x}) X^{{\bar \alpha}_1 \ldots {\bar \alpha}_k}({\bar x})
\end{equation}
which may depend parametrically on the observation point
but it suffices to consider the action of the Green function on
a scalar :
\begin{equation}
- \frac{1}{\Box}X^{\alpha_1 \ldots \alpha_k}(x) =
\frac{1}{4\pi} \int\limits_{{past\: of}\; x} d{\bar x} {\bar g}^{1/2}
\delta ( \sigma( x, {\bar x} )){\cal Y}({\bar x}) + O[X \times \Re] .
\end{equation}
The integration over the light cone in (B.6) includes subintegrations
along the light rays coming from ${\cal I}^-$ to the observation
point $x$ :
\begin{equation}
\int\limits_0^{\infty} d\mu (\rho) {\cal Y}\Bigl|_{L}\; .
\end{equation}
Here  $L$ is a generator of the past light cone of $x, \rho$
 is the luminosity parameter along $L,$ and the measure in
(B.7) is asymptotically of the form
\begin{equation}
d \mu (\rho)\Bigl|_{\rho \to \infty} = d \rho \cdot \rho\; .
\end{equation}
We shall, therefore, assume that the source decreases at
${\cal I}^-$ like
\begin{equation}
X^{\alpha_1 \ldots \alpha_k}\Bigl|_{{\cal I}^-} =
O\left( \frac{1}{r^3}\right)\; .
\end{equation}
Another important assumption [6] is analyticity of the source
in time including the past timelike infinity $(i^-).$ The real
sources appearing in the expectation-value equations are built
out of the curvature, and the condition of analyticity implies
in particular that, at $i^-,$ the metric becomes asymptotically
static. This should be provided by imposing the respective
requirement on $T^{\mu\nu}_{source}.$ By analyticity, the
 limit $r \to \infty$ of the source at $i^-$ coincides with its
limit in the past of ${\cal I}^-:$
\begin{equation}
\Bigl(X^{\alpha_1 \ldots \alpha_k}
\Bigl|_{i^-}\Bigr)\biggr|_{r \to \infty} =
\Bigl(X^{\alpha_1 \ldots \alpha_k}
\Bigl|_{{\cal I}^-}\Bigr)\biggr|_{v \to -\infty}
\end{equation}
where $v$ is the advanced time along ${\cal I}^-.$ Hence,
by (B.9),
\begin{equation}
\Bigl(X^{\alpha_1 \ldots \alpha_k}
\Bigl|_{i^-}\Bigr)\biggr|_{r \to \infty} = O\left(
\frac{1}{r^3}\right)\; .
\end{equation}

For the calculation of the integral (B.6) at $x \to {\cal I}^+$
we may use the Bondi-Sachs frame (A.1). For the past of
${\cal I}^+$ this is safe even if the metric  has closed apparent
horizons [2,3] since no one of these will be encountered by
 the light rays emitted sufficiently early. To lowest order in
the curvature, the world function is then of the form
\begin{equation}
\sigma( x, {\bar x}) = - (u - {\bar u})
( r - {\bar r} + \frac{u - {\bar u}}{2}) +
r{\bar r}(1 - \cos \omega) + O[R_{..}]
\end{equation}
where $\frac{1}{2}\omega^2 $ is the world function on the
2-sphere :
\begin{equation}
\cos \omega = \cos \theta \cos {\bar \theta} +
\sin \theta \sin {\bar \theta} \cos (\varphi -
{\bar \varphi}) \; .
\end{equation}
By solving the equation $ \sigma (x, {\bar x}) = 0$ with
respect to ${\bar u}$ and choosing the solution which
corresponds to the past light cone of $x, {\bar u} = f,$
we obtain
\begin{eqnarray}
-\frac{1}{\Box} X^{\alpha_1 \ldots \alpha_k}
( r, \theta, \varphi, u ) = \frac{1}{4\pi} \int
d^2 {\bar {\cal S}} \int\limits^{\infty}_{0}
d{\bar r} {\bar r}^2 \Bigl| ({\bar \nabla}
{\bar u}, {\bar \nabla}{\bar r}) \cdot
\frac{\partial \sigma}{\partial {\bar u}}
\Bigr|^{-1} {\cal Y}({\bar r}, {\bar \theta}, {\bar \varphi},
{\bar u})\biggl|_{{\bar u} = f} + \nonumber \\
+\; O[X\times\Re] \qquad
\end{eqnarray}
and
\begin{equation}
-\frac{1}{\Box} X^{\alpha_1 \ldots \alpha_k}
( r, \theta, \varphi, u )\biggr|_{{\cal I}^+} =
\frac{1}{r}\: Q(\theta, \varphi, u) + O\left(\frac{1}{r^2}\right)
\end{equation}
where
\begin{equation}
Q( \theta, \varphi, u) = - \frac{1}{4\pi} \int d^2
{\bar {\cal S}} \int\limits_{0}^{\infty} d {\bar r}
(\log{\bar r}) \frac{\partial}{\partial {\bar r}}
\Bigl[ {\bar r}^3 {\cal Y}({\bar r}, {\bar \theta},
{\bar \varphi}, {\bar u})\Bigl|_{{\bar u} = f^{*}}
\Bigr]
+\; O[ X\times\Re ] \; ,\quad
\end{equation}
\begin{equation}
f^{*} = u - {\bar r}(1 - \cos \omega) + O[R_{..}]\;,
\end{equation}
and the equation ${\bar u}=f^{*}$ is the equation of the
limiting light cone of the point $x$ at ${\cal I}^+.$ Here we
wrote $ {\bar r}^2 = {\bar r}^3 {\partial \log {\bar
r}}/{\partial {\bar r}}$ and integrated by parts for being
able to consider sources decreasing at ${\cal I}^-$ like
$1/r^3.$ At the limit $u \to -\infty,$ the source in (B.16)
turns out to be at ${\bar u} \to -\infty.$ By the assumption
of analyticity, we then have
\begin{equation}
Q( \theta, \varphi, u )\Bigl|_{u \to -\infty} =
Q_0 + \frac{1}{u}Q_1 + \cdots
\end{equation}
with
\begin{equation}
Q_0 = - \frac{1}{4\pi} \int d^2
{\bar {\cal S}} \int\limits_{0}^{\infty} d {\bar r}
(\log{\bar r}) \frac{\partial}{\partial {\bar r}}
\Bigl[ {\bar r}^3 {\cal Y}({\bar r}, {\bar \theta},
{\bar \varphi}, -\infty)\Bigr] +  O[ X\times\Re ] \; ,
\end{equation}
etc., and the convergence of the integral in (B.19) is
now owing to (B.11).

Eqs. (B.15)-(B.19) make manifest the fact which is more
general than the approximations made. Namely, the integral
(B.1) in the past of ${\cal I}^+$ involves only the source
$X^{\alpha_1 \ldots \alpha_k}$ at $i^-.$ Under the
assumption of analyticity at $i^-,$ this integral becomes a
static Coulomb potential. One can show that it preserves
this form also at two other limits : at the past null infinity
and spatial infinity ($i^0$),
\begin{equation}
- \frac{1}{\Box} X^{\alpha_1 \ldots \alpha_k}\biggl|_{{\cal
I}^- or\: i^0\: or {\cal I}^+, u \to -\infty} =
\frac{Q_0}{r} (1 + {\cal O})\;,\quad {\cal O} \to 0 \;.
\end{equation}

Consider now equation (5.9) for the news functions which is
valid under the assumption made in sec. 2 that the flux
 components of the Ricci tensor at ${\cal I}^-$ vanish.
Since, in this case, $R^{\mu\nu}$ at ${\cal I}^-$ is
$O(1/r^3),$ the integral implied in $(1/\Box) R^{\mu\nu}$
converges. Moreover, the derivatives of this integral in the
null-tetrad basis behave at ${\cal I}^-$ like
\begin{equation}
\Bigl( \nabla_{\gamma} \frac{1}{\Box} R^{\mu\nu}
\Bigr) \biggr|_{{\cal I}^-} = O\left( \frac{1}{r^2}\right)
\end{equation}
(see below). Therefore, the nonlinear additions to
 $R^{\nu\beta}$ in eq. (5.10) are $O(1/r^4)$ at
${\cal I}^-.$ We then have
\footnote{ Since, at ${\cal I}^+$, the flux components of
$R^{\mu\nu}$ do not vanish, the integral $(1/\Box)
R^{\mu\nu}$ behaves at ${\cal I}^+$ like $(\log r)/r .$
The $(\log r)/r$ asymptotic terms vanish in expression
(2.4b) owing to the presence of the antisymmetrized
derivatives, and in expressions (5.10), (5.13), (B.22)
owing to the contraction with $m_{\beta}m_{\nu}.$}
\begin{eqnarray}
m_{\beta}m_{\nu} \frac{1}{\Box} \Bigl[ R^{\nu\beta}
+ (\nabla^{\nu} \frac{1}{\Box} R^{\gamma\delta})(
\nabla^{\beta} \frac{1}{\Box} R_{\gamma\delta}) -
2 (\nabla_{\gamma} \frac{1}{\Box} R^{\nu\delta})(
\nabla_{\delta} \frac{1}{\Box} R^{\beta\gamma}) +
 O[ R^3_{..}] \Bigr]\biggr|_{{\cal I}^+} =
 \nonumber \\
 {} = \frac{1}{r} q( \theta, \varphi, u) + O\left( \frac{1}{r^2}\right)
\; .\qquad\quad
\end{eqnarray}
We need to calculate the limit
\begin{equation}
\lim_{u \to -\infty} \frac{\partial}{\partial u}\; q( \theta, \varphi,
u)
\end{equation}
which serves as an initial datum for eq. (5.9). By analyticity
of the metric at $i^-,$
\begin{equation}
q( \theta, \varphi, u)\Bigl|_{u \to -\infty} = q_0 +
\frac{1}{u}\: q_1 + \cdots \;,
\end{equation}
and the limit (B.23) vanishes owing to the presence of the
 derivative ${\partial}/{\partial u}.$

Eq. (B.20) can also be used to prove that the solution of the
 Bianchi identities (2.3) with zero initial data for the
gravitational field at ${\cal I}^-$ is expressed indeed in terms
 of the retarded Green function. This is easily seen in expression
(2.4b). Since, by (B.20),
 \begin{equation}
- \frac{1}{\Box}\Bigl( R^{\nu\beta} +  O[R^2_{..}]
\Bigr) \biggr|_{{\cal I}^-} =
\frac{Q^{\nu\beta}_0}{r} + O\left(\frac{1}{r^2}\right) \; ,
\end{equation}
and $Q^{\nu\beta}_0$ {\it does not depend on the advanced
time along} ${\cal I}^-,$ the derivatives in (2.4b) enhance
the power of $1/r.$ Eq. (B.21) is valid for the same reason.

The proof is more involved in the general case, where the fluxes
of $R^{\mu\nu}$ through ${\cal I}^-$ are nonvanishing,
since one has to address eq. (2.4a). In terms of the null tetrad
 at ${\cal I}^-$
\begin{equation}
\nabla^{[\mu} \nabla^{<\alpha} R^{\nu]\beta>}\biggl|_{
{\cal I}^-} = \frac{\nabla^{[\mu}v \nabla^{<\alpha}v}{
(\nabla u, \nabla v)^2} (\nabla_{\gamma}u) ( \nabla_{\sigma}
u) \nabla^{\gamma}\nabla^{\sigma} R^{\nu]\beta>} +
O\left( \frac{1}{r^3}\right) \; .
\end{equation}
Since the flux components of $R^{\nu\beta}$ at ${\cal I}^-$
are proportional either to $\nabla^{\nu}v$ or to $\nabla^{\beta}v,$
they cancel in (B.26)  owing to the antisymmetrizations :
\begin{equation}
\nabla^{[\mu} \nabla^{<\alpha} R^{\nu]\beta>}\Bigl|_{
{\cal I}^-} = O\left( \frac{1}{r^3}\right)\; .
\end{equation}
Thus the source in (2.4a) satisfies condition (B.9), and we have
\begin{equation}
R^{\alpha\beta\mu\nu}\biggl|_{{\cal I}^-} =
- \frac{Q_0^{\alpha\beta\mu\nu}}{r} + O\left(\frac{1}{r^2}\right)
\end{equation}
with $Q_0^{\alpha\beta\mu\nu}$ constant along ${\cal I}^-.$
By (B.19) and (B.5),
\begin{eqnarray}
Q_0^{\alpha\beta\mu\nu}& =& - \frac{1}{4\pi} \int d^2 {\bar {
\cal S}} \int\limits_0^{\infty} d{\bar r} (\log {\bar r})
\frac{\partial}{\partial {\bar r}} \left( {\bar r}^3
{\bar {\cal Y}}\Bigl|_{i^-}\right) + O[R^2_{..}] \; , \\
{\bar {\cal Y}}& =& 4 g^{[\mu}_{\;\:{\bar \mu}}
g^{\nu]}_{\;\:{\bar \nu}} g^{[\alpha}_{\;\:{\bar \alpha}}
g^{\beta]}_{\;\:{\bar \beta}} {\bar \nabla}^{\bar \mu}
{\bar \nabla}^{\bar \alpha} {\bar R}^{{\bar \nu}{\bar \beta}}\; ,
\end{eqnarray}
and, up to $O[R^2_{..}], {\bar {\cal Y}}$ is a total derivative:
\begin{eqnarray}
{\bar {\cal Y}}\biggl|_{i^-}& =& {\bar g}^{-1/2}
{\bar \partial}_{\bar \mu} \left( {\bar g}^{1/2} {\bar Z}^{\bar \mu}
\right) \biggl|_{i^-} +\;\: O[R^2_{..}]\; , \\
{\bar Z}^{\bar \mu}& =& 4 g^{[\mu{\bar \mu}}g^{\nu]{\bar \nu}}
 g^{[\alpha{\bar \alpha}}g^{\beta]{\bar \beta}}
{\bar \nabla}_{\bar \alpha} {\bar R}_{{\bar \nu}{\bar \beta}} \; ,
\end{eqnarray}
where use is made of eq. (B.3). Since, in this total derivative,
the time derivative vanishes owing to the asymptotic
stationarity of the metric at $i^-,$ and the angle derivatives
vanish in the integral $\int d^2{\bar {\cal S}},$ we obtain
\begin{equation}
Q_0^{\alpha\beta\mu\nu}  =  - \frac{1}{4\pi} \int d^2 {\bar {
\cal S}} \int\limits_0^{\infty} d{\bar r} (\log {\bar r})
\frac{\partial}{\partial {\bar r}} \left[ {\bar r}
\frac{\partial}{\partial {\bar r}} \Bigl( {\bar r}^2
{\bar \nabla}_{\bar \mu} {\bar r}\; {\bar Z}^{\bar \mu}
\biggl|_{i^-}\Bigr)\right]  + O[R^2_{..}]\; .
\end{equation}
Since the power of decrease of ${\bar Z}^{\bar \mu}\Bigl|_{i^-}$
at ${\bar r} \to \infty $ is at least $O(1/{{\bar r}^2}),$ we may
integrate by parts in (B.33) {\it to remove the integration} over
 ${\bar r}$ completely :
\begin{equation}
Q_0^{\alpha\beta\mu\nu}  = \lim_{{\bar r} \to \infty}
 \frac{1}{4\pi} \int d^2 {\bar {
\cal S}}\Bigl( 4 g^{[\mu{\bar \mu}}g^{\nu]{\bar \nu}}
 g^{[\alpha{\bar \alpha}}g^{\beta]{\bar \beta}}
{\bar \nabla}_{\bar \mu}{\bar r} \Bigr)
\Bigl({\bar r}^2 {\bar \nabla}_{\bar \alpha}
{\bar R}_{{\bar \nu}{\bar \beta}}\biggl|_{i^-}\Bigr)
+ O[R^2_{..}] \; .
\end{equation}
Only the terms in ${\bar R}_{{\bar \nu}{\bar \beta}}$ that
decrease like $1/{{\bar r}^2}$ can survive in (B.34).
However, the integrand in (B.34) contains one more
derivative of ${\bar R}_{{\bar \nu}{\bar \beta}},$ and,
again, the respective time derivative vanishes at $i^-.$
Therefore, this integrand is $O(1/{\bar r})$
(actually $O(1/{{\bar r}^2}),$ see below), and the limit
(B.34) vanishes. The presence of an extra derivative is in
fact not essential. By (B.10), the sequence of limits in
(B.34) can be replaced by the limit of the same
quantity in the past of ${\cal I}^-.$ It is then seen that
the constant (B.34) hangs solely on the flux components
of the Ricci tensor {\it in the past of} ${\cal I}^-.$
However, the metric in the past becomes asymptotically
static, and for a static metric the fluxes of $R_{\nu\beta}$
through ${\cal I}^-$ are absent as seen from the counterparts
 at ${\cal I}^-$ of eqs. (A.18) - (A.21). We have, therefore,
\begin{equation}
( R_{\nu\beta}\Bigl|_{i^-} )\biggl|_{r \to \infty} =
( R_{\nu\beta}\Bigl|_{{\cal I}^-} )\biggl|_{v \to -\infty} =
O\left(\frac{1}{r^3}\right)\; ,
\end{equation}
and
\begin{equation}
Q_0^{\alpha\beta\mu\nu} = 0\; .
\end{equation}
Hence
\begin{equation}
R^{\alpha\beta\mu\nu}\Bigl|_{{\cal I}^-} = O\left(\frac{1}{r^2}
\right)
\end{equation}
which proves that, in the retarded solution, there are no incoming
gravitational waves.
    \section*{References.}
   \begin{enumerate}
\item  V.P.Frolov and G.A.Vilkovisky, \underline{ Proc. 2nd Marcel
Grossmann Meeting on} \\
\underline{ General Relativity (Trieste 1979)}, ed.
R.Ruffini (Amsterdam: North-Holland 1982) p.455 \\
\item V.P.Frolov and G.A.Vilkovisky, Phys. Lett. B 106 (1981) 307 \\
\item V.P.Frolov and G.A.Vilkovisky,
\underline{Proc. 2nd Seminar on Quantum Gravity}\\
\underline{(Moscow 1981)},
 ed. M. A. Markov and P.C.West (London: Plenum, 1983) p.~267 \\
\item  G.A.Vilkovisky,  Class. Quantum Grav. 9 (1992) 895 \\
\item A.O.Barvinsky, Yu.V.Gusev, G.A.Vilkovisky, and
V.V.Zhytnikov, J.Math. Phys. 35 (1994) 3525 \\
\item A.G.Mirzabekian and G.A.Vilkovisky, Phys. Lett. B 317
(1993) 517 \\
\item A.G.Mirzabekian, Zh. Eksp. Teor. Fiz. 106 (1994) 5
[ JETP 79 (1994) 1 ] \\
\item A.O.Barvinsky  and G.A.Vilkovisky, Nucl. Phys. B 282 (1987) 163\\
\item A.O.Barvinsky  and G.A.Vilkovisky, Nucl. Phys. B 333 (1990) 471\\
\item A.O.Barvinsky  and G.A.Vilkovisky, Nucl. Phys. B 333 (1990) 512\\
\item G.A.Vilkovisky, Preprint CERN-TH. 6392/92;
\underline{Publication de l`Institut de} \\
\underline{Recherche Mathematique Avancee, R.C.P. 25}
, vol. 43 (Strasbourg, 1992) p.203\\
\item A.O.Barvinsky, Yu.V.Gusev, G.A.Vilkovisky,and V.V.Zhytnikov,
\underline{Covariant}\\
\underline{Perturbation Theory ( IV), Third Order in The
Curvature}, Report of the University of Manitoba ( Winnipeg :
U. of Manitoba 1993) pp. 1-192\\
\item A.O.Barvinsky, Yu.V.Gusev, G.A.Vilkovisky, and
V.V.Zhytnikov, J.Math. Phys. 35 (1994) 3543\\
\item A.O.Barvinsky, Yu.V.Gusev, G.A.Vilkovisky, and
V.V.Zhytnikov, Nucl. Phys. B, to appear\\
\item A.O.Barvinsky, Yu.V.Gusev, V.V.Zhytnikov, and
G.A.Vilkovisky, \underline{ Asymptotic}
\underline{behaviours of
the one-loop vertices in the gravitational effective action}
(to be published)\\
\item L.P.Grishchuk, Zh. Eksp. Teor. Fiz. 67 (1974) 825
[ Sov. Phys. JETP 40 (1975) 409 ]\\
\item L.P.Grishchuk and Yu.V.Sidorov, Phys. Rev. D 42 (1990)
3413\\
\item L.P.Grishchuk, Class. Quantum Grav. 10 (1993) 2449\\
\item B.S.DeWitt, \underline{Relativity, groups and topology
(1983 Les Houches lectures)}, ed. B.S.DeWitt and R.Stora
(Amsterdam:North-Holland, 1984) p. 221\\
\item H.Bondi, M.G.J. van der Burg, and A.W.K.Metzner,
Proc. R. Soc. London A269 (1962) 21\\
\item R.Sachs, \underline{Relativity, groups and topology
( 1963 Les Houches lectures)}, ed. C.DeWitt and
B.S.DeWitt ( New York: Gordon and Breach 1964)\\
\item R.M.Wald, \underline{General Relativity} (Chicago: The
University of Chicago Press, 1984)\\
\item J.L.Synge, \underline{ Relativity: The General Theory }
(Amsterdam: North-Holland 1960~)\\
\item B.S.DeWitt,
\underline{Dynamical Theory of Groups and Fields}
 (New York: Gordon and Breach, 1965)
\end{enumerate}
\end{document}